\documentclass[prb,twocolumn,floatfix,amsfonts]{revtex4-1}
\usepackage{amsmath}
\usepackage{bm}
\usepackage{color}
\usepackage{graphicx,graphics,color,epsfig}% Include figure files
\usepackage{amssymb}

\graphicspath{{./Figures/}}

\begin{document}

% Title of the article
\title{Comparative Scanning Tunneling Microscopy Study on Hexaborides}

\author{Steffen Wirth}\email{steffen.wirth@cpfs.mpg.de}
\affiliation{Max Planck Institute for Chemical Physics of Solids, 01187
Dresden, Germany}

\author{Sahana R\"{o}{\ss}ler}
\affiliation{Max Planck Institute for Chemical Physics of Solids, 01187
Dresden, Germany}

\author{Lin Jiao}
\email{Present address: Department of Physics and Frederick
   Seitz Materials Research Laboratory, University of Illinois
   Urbana-Champaign, Urbana, Illinois 61801, USA}
\affiliation{Max Planck Institute for Chemical Physics of Solids, 01187
Dresden, Germany}

\author{M. Victoria Ale Crivillero}
\affiliation{Max Planck Institute for Chemical Physics of Solids, 01187
Dresden, Germany}

\author{Priscila F. S. Rosa}
\affiliation{Los Alamos National Laboratory, Los Alamos, NM 87545, USA}

\author{Zachary Fisk}
\affiliation{Department of Physics, University of California,
  Irvine, California 92697, USA}

\date{\today}

\begin{abstract}
%  The hexaborides SmB$_6$ and EuB$_6$ cover an intriguing range of current
%  topical interest including Kondo physics and likely nontrivial surface
%  states in the former and semimetallic behavior and polaron formation in
%  the latter.
We compare STM investigations on two hexaboride compounds, SmB$_6$ and
EuB$_6$, in an effort to provide a comprehensive picture of their surface
structural properties. The latter is of particular importance for studying
the nature of the surface states in SmB$_6$ by surface-sensitive tools.
Beyond the often encountered atomically rough surface topographies of {\it in
situ}, low-temperature cleaved samples, differently reconstructed as well as
B-terminated and, more rarely, rare-earth terminated areas could be found.
With all the different surface topographies observed on both hexaborides, a
reliable assignment of the surface terminations can be brought forward.
\end{abstract}

% Please select about four verbal keywords for your manuscript.
\keywords{hexaborides, scanning tunneling microscopy and spectroscopy, surface
terminations}

\maketitle   % please do not remove

\section{Introduction}

The hexaborides of cubic structural type CaB$_6$ ($Pm\bar{3}m$) represent a
very versatile class of compounds \cite{eto85}. LaB$_6$ features a very low
work function of about 2.7 eV \cite{ber78} while electron-doped CaB$_6$ is
a ferromagnetic material, albeit with low magnetic moment \cite{you99,sta14},
and CeB$_6$ exhibits quadrupolar ordering \cite{eff85}. The hexaborides are
often highly conductive. From Hall measurements it was shown \cite{gru85}
that the majority of the hexaborides has one charge carrier per rare-earth
atom, with the exceptions of divalent Eu and Yb with very low charge carrier
densities and SmB$_6$ exhibiting intermediate valence \cite{vai64,uts17}.

The material SmB$_6$ has attracted special attention recently as it was
proposed to host topologically non-trivial surface states \cite{dze10}. This
material falls into the category of so-called Kondo insulators
\cite{aep92,ris00} in which the insulating properties are brought about by
hybridization between conduction bands (here $d$-bands) and localized
$f$-states. In consequence, a narrow gap opens up at sufficiently low
temperatures (below the Kondo temperature $T_{\rm K}$) while the
$f$-electrons provide the strong spin-orbit coupling required for the
development of topologically protected surface states predicted by band
structure calculations \cite{tak11,lu13,jkim14}. Subsequently, considerable
experimental effort was made to verify the topological nature of the surface
states, in particular through angle-resolved photoemission spectroscopy
(ARPES) with spin resolution \cite{nxu14,sug14}. Though there is a consensus
on the existence of a conducting surface state \cite{wol13,eo19}, its
topological nature is a matter of ongoing debate. For instance, the
surface states observed by ARPES measurements have been interpreted in
terms of Rashba splitting \cite{hla18}. One crucial aspect \cite{bar15,leg15},
namely a $\Gamma_8$ quartet ground state of the Sm $f^5$ configuration, has
recently been observed experimentally \cite{sev18}, but is in contrast to
some band structure calculations \cite{lu13,ant02,kan15}. Here, the strong
correlations of the Kondo insulator as well as its intermediate valence
complicate band structure calculations \cite{min17}. An additional
complication is the complex $(001)$ surface of SmB$_6$ itself due to its
polar nature \cite{zhu13}. Because of the cubic structure of SmB$_6$, {\it
in situ} cleaved surfaces usually required for surface-sensitive techniques
like ARPES or Scanning Tunneling Microscopy/Spectroscopy (STM/S) are often
atomically rough or reconstructed \cite{ruan14,roe14,sun18,pir19}. But even
in case of atomically flat surfaces the interpretation of the surface
termination in STM is controversial \cite{roe16,her18}. In an effort to make
progress in this complex situation we here compare topographies obtained by
STM on SmB$_6$ and EuB$_6$. The latter material is interesting in its own
right due to its complex band structure \cite{zha08}, ferromagnetic properties \cite{sue98} and polaron formation \cite{poh18}. We note that a comparative
study of YbB$_6$, CeB$_6$ and SmB$_6$, primarily based on ARPES results, was
recently brought forward \cite{ram16}.

\section{Experimental Section}
\label{II}
Single crystals of SmB$_6$ and EuB$_6$ were grown by an Al flux method
\cite{fis79,kim14,rosa18}. The orientation of the single crystals was checked
by Laue diffraction. The lattice constants are $a =$ 4.133 {\AA} for SmB$_6$
and $a =$ 4.185 {\AA} for EuB$_6$.

For STM investigations, two ultra-high vacuum (UHV) systems were used
\cite{omi}. A $^4$He system allows for base temperatures below 5 K; if a
heating stage is used the base temperature is typically $\sim$6 K (a
temperature sensor is incorporated into the heating stage). If not stated
otherwise, the STM/STS data reported in the following were obtained at
\begin{figure}[t]%
\centering
\includegraphics*[width=0.48\textwidth]{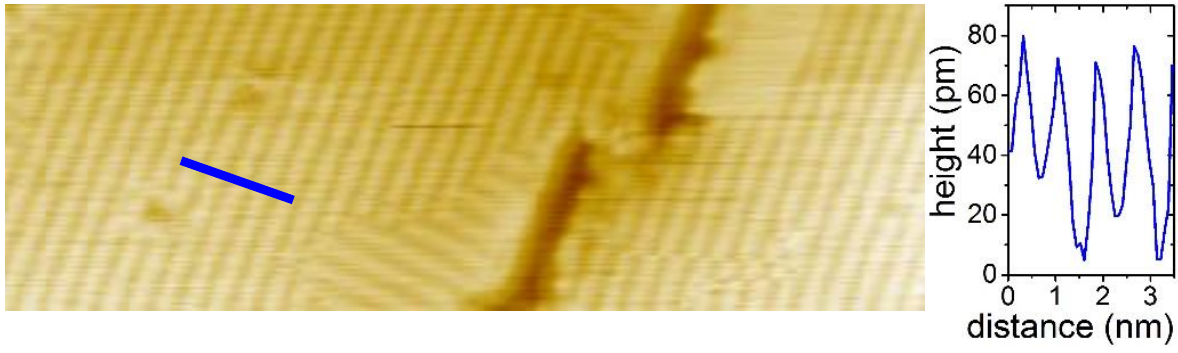}
\caption{Topography (10 nm$\,\times\,$30 nm) of a reconstructed area of
SmB$_6$. Twinning between (2$\,\times\,$1) and (1$\,\times\,$2)
reconstructions and dislocations of the reconstruction can clearly be seen
($V = 0.2$~V, $I_{\rm sp} = 0.5$ nA). The height scan (right) was taken
along the blue line indicated in the topography.}
\label{twin}
\end{figure}
$\sim$6~K. Our $^3$He-based system operates down to a base temperature of
$\sim$0.3 K and allows to apply magnetic fields up to 12 T perpendicular to
the investigated surface. Electrochemically etched tungsten tips were used
if not stated otherwise. Tunneling spectroscopy was conducted by using a
lock-in technique and adding a small ac modulation voltage $V_{\rm mod}$
of 0.1 or 1.0 mV (depending on bias voltage $V$, see respective
figure caption) with a frequency of 117 Hz to the bias voltage.

Some of the STM data reported in the following were obtained in a so-called
dual-bias mode. In these cases, two different bias voltages $V$ were applied
for the forward and backward scan of the fast scan direction. This mode
of operation allows to obtain topographic images with two different $V$ at
the same sample position (within the piezoelectric hysteresis of the scanner,
typically giving offsets well below 1\% of the total scan size of the two
topographies). In doing so, drift corrections can be neglected and parameters
like temperature $T$ or magnetic field, sample history, surface termination,
or tip condition are identical. Even if the tip changes, it then
influences the data for both $V$ at very similar sample positions.

All samples reported here were cleaved {\it in situ} along a $(001)$
crystallographic plane at a temperature of $\sim$20~K using identical cleaving
stages in both UHV systems. After cleaving, the sample needs to be transferred
into the respective STM head during which time (in the order of 10 s) the
sample temperature is not controlled. We here provide results based on 24
cleaves of SmB$_6$ (on 8 of which we did not find any atomically flat surface
area) and 5 cleaves on EuB$_6$.

\section{Results}
\subsection{SmB$_6$}
In order to obtain information on the nature of the surface states the
applied probe needs to be surface sensitive. One obstacle in investigating
SmB$_6$ with highly surface-sensitive tools like ARPES or STM are the
\begin{figure}[t]%
\centering
\includegraphics*[width=0.48\textwidth]{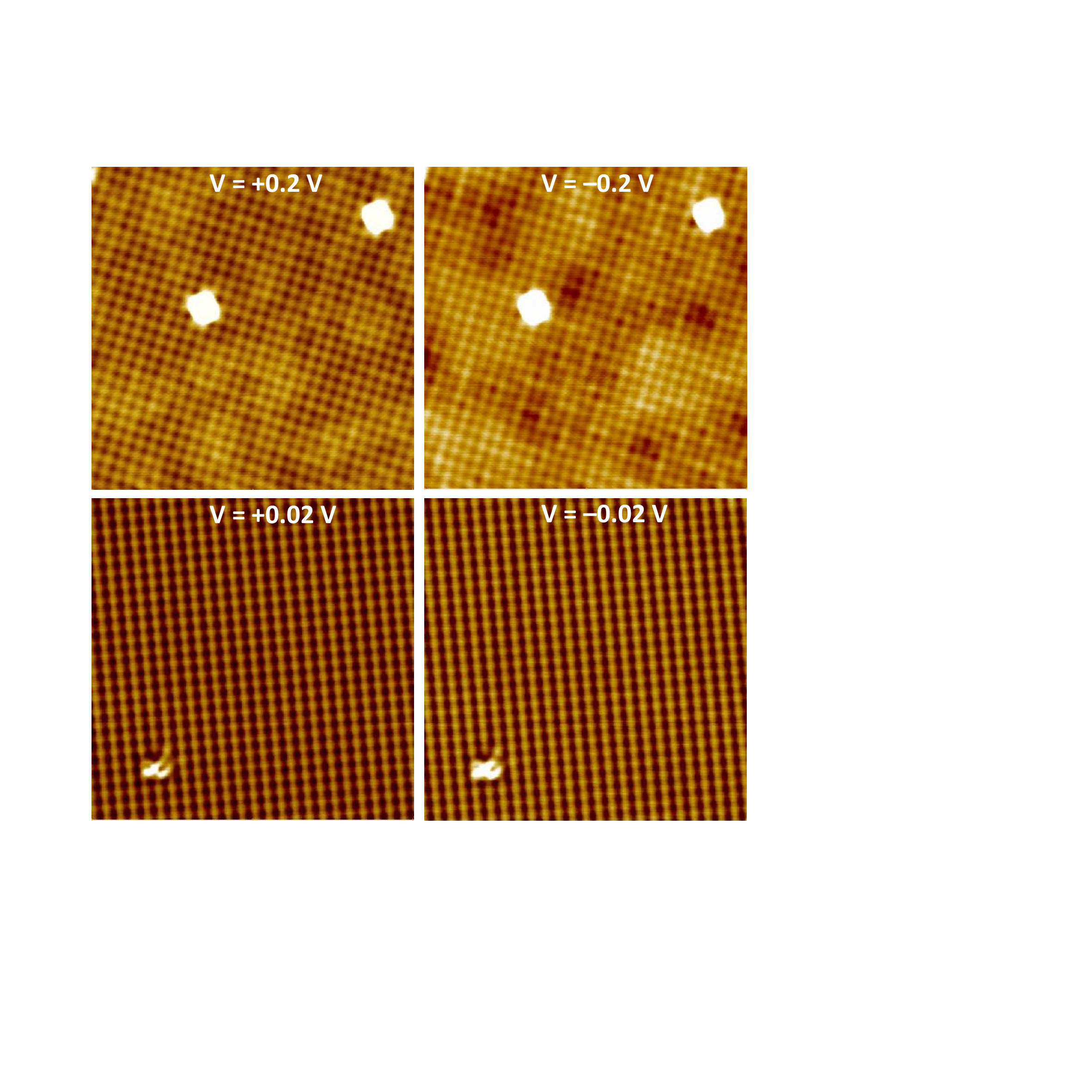}
\caption{The upper and lower row show two different SmB$_6$ topographies
of areas 10 nm$\,\times\,$10 nm, both obtained in dual-bias mode: upper
row $V =+0.2$ V (left) and $V = -0.2$ V (right), $I_{\rm sp} =$ 0.6 nA; lower
row $V =+0.02$ V (left) and $V = -0.02$ V (right), $I_{\rm sp} =$ 0.3 nA.
Images in the upper (lower) row were obtained at $\sim$6 K ($\sim$1.7 K) and
cover a total height scale of 155 pm (90 pm). Also the defects indicate that
exactly the same sample areas are imaged on the left and right side,
respectively.}  \label{topo-def}
\end{figure}
different surface terminations. Due to the cubic structure of the hexaborides,
the majority of the cleaved surface areas is rough on an atomic scale
\cite{roe16}. This may result in a modified local structure which, in turn,
may influence the properties (specifically of Sm) at the surface
\cite{schu12,wol15}.

Upon searching, a $(2\,\times\,1)$ surface reconstruction can usually be
found \cite{roe14,pir19,yee13}. The $(2\,\times\,1)$ reconstruction was
also observed by low-energy electron diffraction \cite{ram16,miy12,miy17}
as well as by STM on LaB$_6$ \cite{buc19}. Clearly, if we assume that the
$(2\,\times\,1)$ reconstruction results from each second row of Sm atoms
\begin{figure*}[t]%
\centering
\includegraphics*[width=0.8\textwidth]{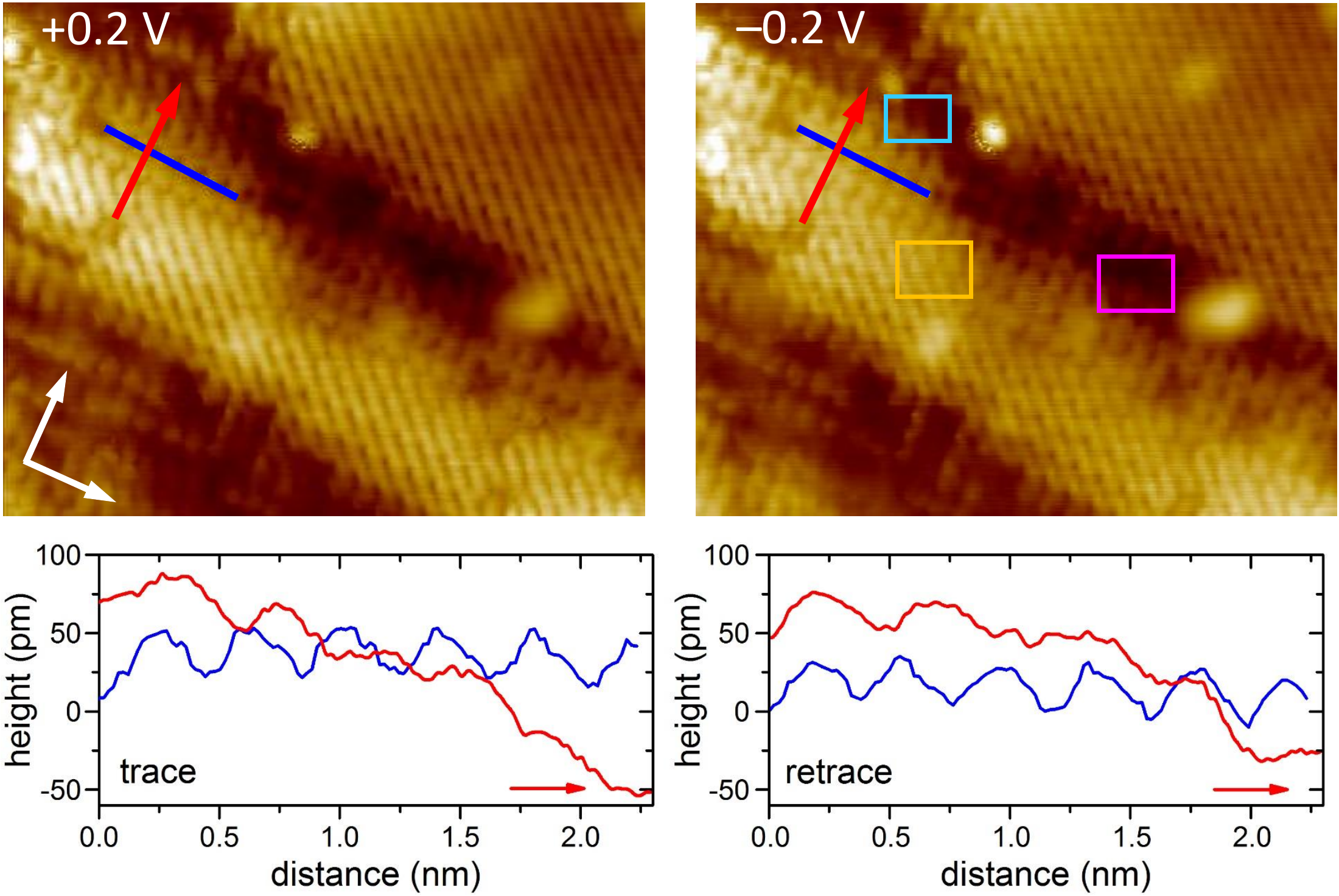}
\caption{SmB$_6$ topography of an area of 10 nm$\,\times\,$8 nm exhibiting
both Sm- and B-terminated areas. The images were obtained in dual-bias mode:
$V =+0.2$ V (left) and $V = -0.2$ V (right). The height scans were taken along
the lines of corresponding color in the topography images. Along the direction
of the red arrows the sample height decreases by about 100 -- 130 pm. The
white arrows indicate the main crystallographic directions $\langle 100
\rangle$ and $\langle 010 \rangle$, respectively. The colored rectangles
indicate the areas over which the spectroscopy curves of corresponding
colors in Fig.\ \ref{trench-spec} were averaged.}
\label{trench}
\end{figure*}
missing on top of an otherwise B-terminated surface, such a reconstruction
is energetically favorable compared to an unreconstructed polar surface.
Yet, other STM studies did not report this reconstruction \cite{sun18} or
interpreted it differently \cite{her18}. It should also be noted that such
a reconstruction may have repercussions on the metallic surface state
\cite{yoo02}.

In Fig.\ \ref{twin} we present such a (2$\,\times\,$1) reconstructed surface
area. The height scan taken along the blue line as indicated in the topography
is consistent with the aforementioned idea of each second row of Sm atoms
missing. This is corroborated by a change in height on and between these rows
of atoms of the order 40--50 pm; yet, an order of magnitude smaller height
oscillations for the (2$\,\times\,$1) reconstructed surface was also reported
\cite{yee13,pir19}. The reconstruction is likely formed during the cleaving
process or subsequently upon some additional diffusion of surface atoms. In
both cases, one may expect domains of (2$\,\times\,$1) and (1$\,\times\,$2)
reconstructed areas and dislocations between the Sm rows by one lattice
constant, both of which can easily be recognized in the topography, Fig.\
\ref{twin}.

Rarely we also found atomically flat surface areas as shown in Fig.\
\ref{topo-def}. Similar surface topographies have been presented before
\cite{ruan14,roe14,sun18,her18,yee13,jiao16,mat18}. It was shown, however,
that the obtained topography depends on applied bias voltage $V$, and even
a contrast reversal was observed for $V =$ 0.2~V and $-$3.0~V \cite{her18}.
In the following we make use the dual-bias mode described in Section \ref{II}
as it allows to visualize exactly the same surface area without relying on
defects on top of the investigated surface (the appearance of defects may
change with $V$, see Fig.\ \ref{topo-def}). We have chosen values of $V$ small
compared to the barrier height $\Phi$ (see also below) yet larger than the
hybridization gap of less than 20 meV
\cite{zha13,nxu13,fra13,fuh17,val18,sun18}. In Fig.\ \ref{topo-def} dual-mode topographies for $V = \pm$ 0.2~V (upper) and $\pm$0.02~V (lower) are compared.
Note that here different samples were investigated at somewhat different
temperatures of $T \sim$ 6~K (upper) and 1.7~K (lower). Qualitatively, the
topographies for given temperature agree very well, i.e.\ there is no contrast
inversion of reversed $V$. There are subtle inhomogeneities in the background
at $T \sim$ 6~K; we can only speculate that they result from a not fully
developed conducting surface state because we so far did not observe such
inhomogeneities at $T \le$ 1.7~K (see also discussion of Fig.\ \ref{defects}
below). We note that very similar inhomogeneities are reported on atomically
flat surfaces, Fig.\ S1B in \cite{pir19}.

To investigate the surface termination in more detail we present in Fig.\
\ref{trench} topographies on areas exhibiting steps of less than one unit cell
height $a$ \cite{roe14}. Such steps are perfectly suited to gain information
about the different surface terminations. Again, the topographies obtained in
dual mode with $V \pm$ 0.2~V agree on a qualitative level. The white arrows
in Fig.\ \ref{trench} indicate the main crystallographic directions $\langle
100 \rangle$ and $\langle 010 \rangle$. Height scans taken along the
blue lines, i.e.\ parallel to one of the main crystallographic directions and
at overall unchanged height, clearly indicate lateral distances between
corrugations consistent with the lattice constant $a$, while such taken along
descending height (red arrows) exhibit less obvious corrugations (possibly
related to crystallographic imperfections within such regions of changing
overall height). Within elevated areas (bright regions), however,
\begin{figure}[t]%
\centering
\includegraphics*[width=0.4\textwidth]{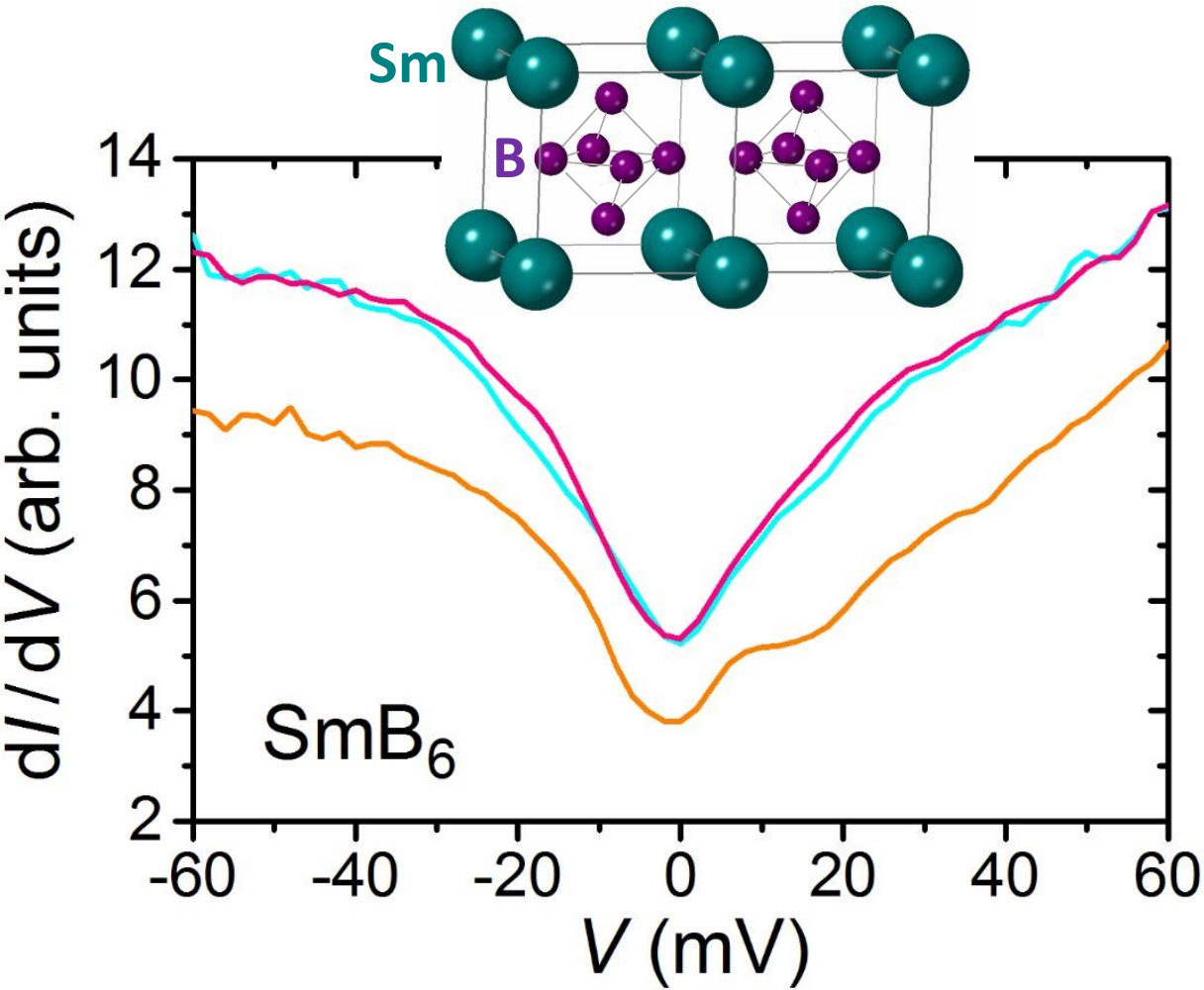}
\caption{Tunneling spectroscopy averaged over the areas marked by rectangles
of corresponding colors in Fig.\ \ref{trench} (right). $V = + 0.2$~V,
$I_{\rm sp} =0.6$~nA, $V_{\rm mod} =1$ mV. The inset exhibits two
adjacent unit cells of the cubic SmB$_6$ lattice.}
\label{trench-spec}
\end{figure}
the corrugations appear to run along the diagonal, i.e.\ $\langle
1\bar{1}0 \rangle$, directions. This is consistent \cite{ruan14,roe14} with a
Sm-terminated surface where also, in addition to the Sm atoms, the apex atoms
of the B-octahedra are seen, see discussion of Fig.\ \ref{compEuSm} below as
well as the crystal structure shown in Fig.\ \ref{trench-spec}.

We now turn to the height scans taken along the red arrows in the topographies
which, again, follow a $\langle 100 \rangle$ direction but also include a
height change. Atomic
\begin{figure}[b]%
  \centering
  \includegraphics[width=.44\textwidth]{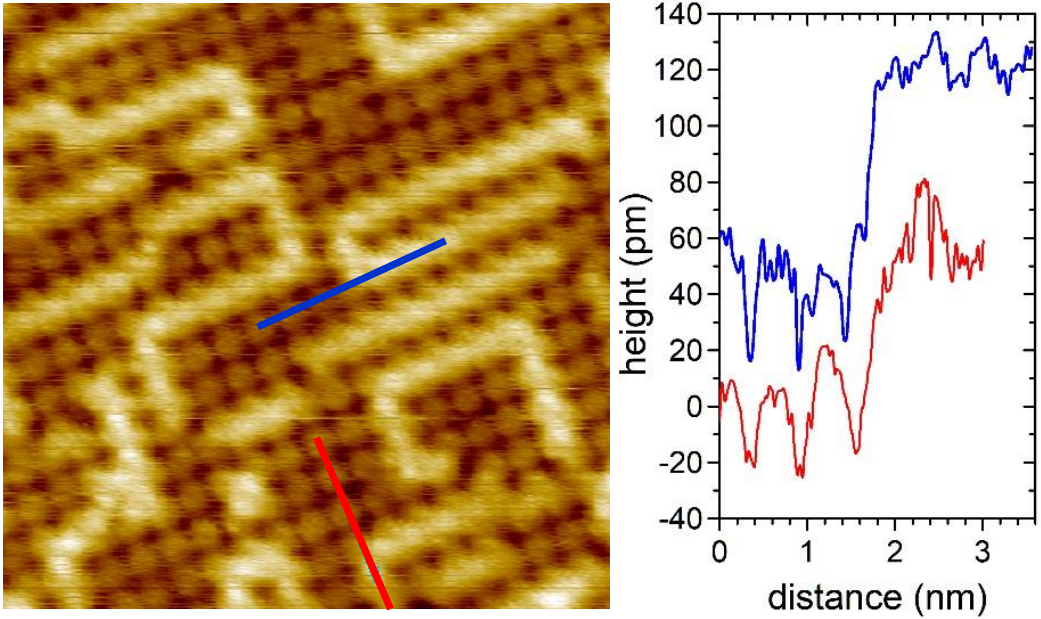}%
  \caption{Disordered reconstructed surface area (10 nm$\,\times\,$10 nm) of
    SmB$_6$. From the height profiles (taken along the lines of
    corresponding colors in the topography) the apparent height of the bright
    atoms is on the order of 60 pm ($V = +0.2$ V, $I_{\rm sp} =$ 0.6 nA,
    blue profile offset by 50 pm).}   \label{recon}
\end{figure}
distances corresponding to $a$ can be seen for $V = +$0.2~V, but less well
for $V = -$0.2~V. Clearly, the total change in height depends to some extent
on $V$: It amounts to about 130 pm for $V = +$0.2~V and $\sim$100 pm for
$-$0.2~V. Yet, both numbers appear to be consistent with the expected step
height upon going from a Sm- to a B-terminated surface considering the
inter-octahedron B distance of 164.6 pm. Given the fact that distances of $a$
are observed along the main crystallographic directions on this $(001)$ plane
such a step height is difficult to interpret otherwise; a viable alternative
is the opposite assignment (i.e. going from a B-terminated surface down to a
Sm-terminated one) which would, however, involve breaking up of B-otcahedra,
i.e. intra-octahedral bond breaking. Estimates of the surface energy
\cite{sun18,roe16} indicate a slight preference for inter-octahedral bond
breaking but impurities or sample inhomogeneities and defects may change
\begin{figure}[t]%
  \centering
  \includegraphics[width=.44\textwidth]{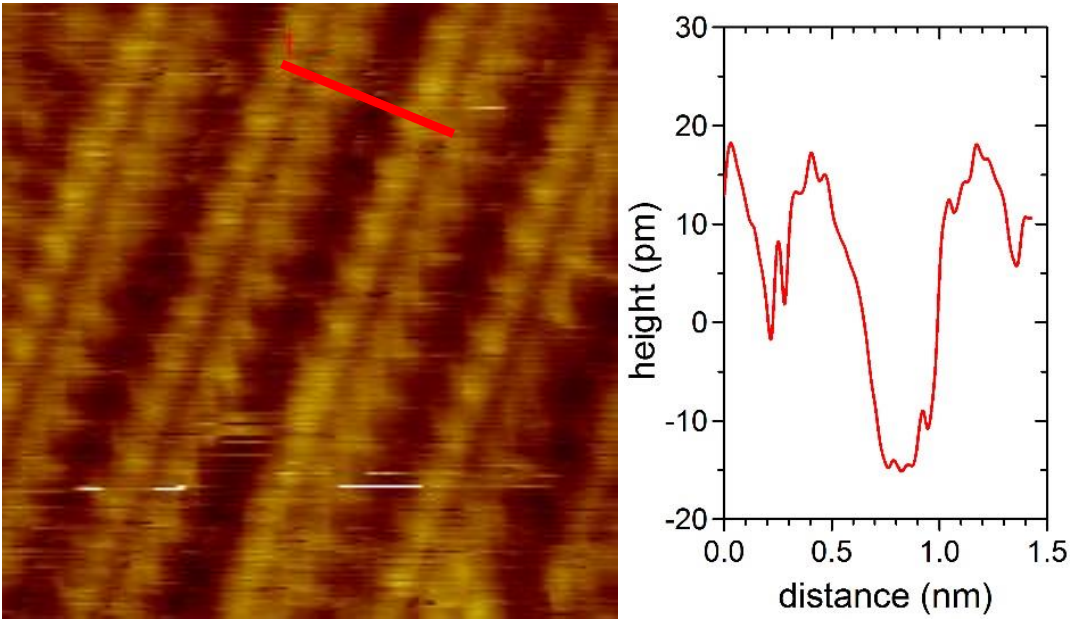}%
  \caption[]{%
    Surface area (5 nm$\,\times\,$5 nm) of SmB$_6$ within which only every
    third row of Sm appears to be missing ($V = +0.2$ V, $I_{\rm sp} = 0.6$
    nA). Right: height scan along the red line shown in the topography.}
    \label{2rows}
\end{figure}
these estimates locally. Indeed, a donut-like structure was interpreted as
breaking inter-octahedral bonds \cite{ruan14}.

In order to gain further insight into the different terminations exposed
in Fig.\ \ref{trench} tunneling spectroscopy was conducted. The STS curves
shown in Fig.\ \ref{trench-spec} correspond in color to the areas marked in
Fig.\ \ref{trench} (right) over which the spectra were averaged. These spectra
can be compared to those obtained on small areas of atomically flat surfaces,
but differ from those seen on larger areas in that there is no pronounced
maximum in d$I$/d$V$ at around $-20$~mV \cite{roe14}. The orange spectrum
attained on the elevated part of the topography Fig. \ \ref{trench} exhibits
a well developed hump at $V \sim +10$ mV \cite{ruan14,roe14,miy17}. It is
tempting to compare this hump to the conspicuous maximum observed on
Sm-terminated surface of larger areas \cite{ruan14,roe14}. Note that we did
not observe (Fig.\ \ref{trench-spec} and \cite{roe14}) a pronounced shift in
energy of features at negative $V$ as reported elsewhere \cite{mat18}.

The discussion above indicated that a $(2\,\times\,1)$ reconstruction is
energetically favorable with respect to the polar nature of a Sm- or
B-terminated surface. However, a similar effect is conceivable if the
$(2\,\times\,1)$ reconstruction is not long-ranged, but realized only locally.
The lines of Sm may then meander \cite{roe14,roe16} not giving rise to a
\begin{figure*}[t]%
\centering
\includegraphics*[width=0.85\textwidth]{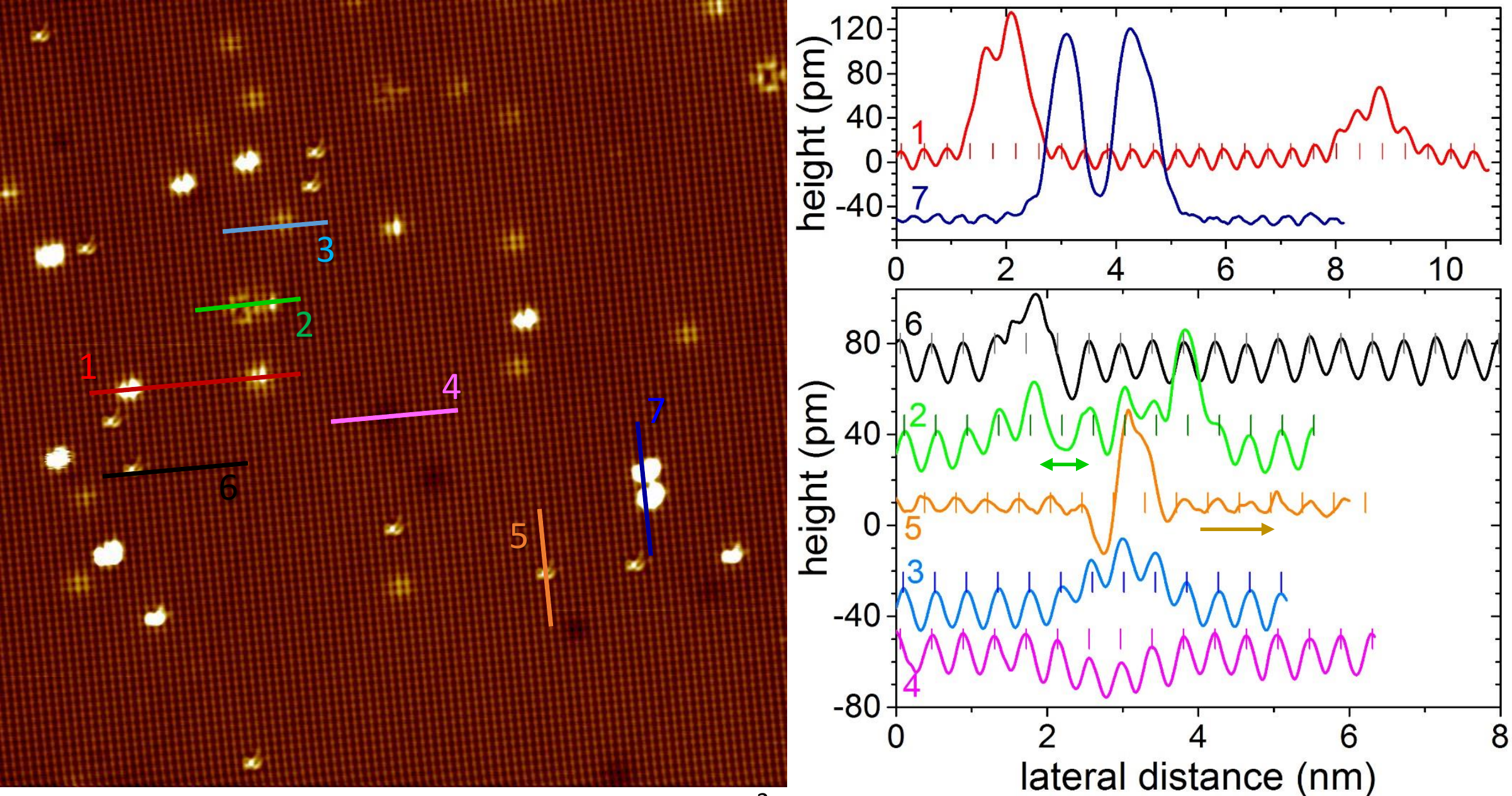}
\caption{Left: SmB$_6$ topography of a B-terminated area of 40 nm$\,\times\,$40
nm with different types of defects ($V =$ 0.015~V, $I_{\rm sp} =$ 0.3~nA, $T
\approx$ 1.7~K). Right: Height scans taken along the lines of corresponding
numbers and colors in the topography image. The dashes mark distances of one
lattice parameter to indicate the expected atomic positions.} \label{defects}
\end{figure*}
superstructure \cite{fra13}. Part of such a ``disordered'' reconstructed
surface is shown in Fig.\ \ref{recon}. In such a case, a similar change in
height upon going from the topmost Sm atoms to the underlying B layer is
expected as in Fig.\ \ref{trench}. The height scan, Fig.\ \ref{recon} (right),
along the red line marked in the topography indeed supports this assertion.

In one instance, we observed a topography as presented in Fig.\ \ref{2rows}.
The height scan may be interpreted as every third row of atoms missing. Here,
the height change between the upper and lower rows of atoms is only about
30 pm, similar to \cite{roe14} or slightly smaller (Fig.\ \ref{twin}) than
the case of $(2\,\times\,1)$ reconstructions. The exact number, however, may
depend on details of the tip, i.e. how well it may penetrate between the rows
of atoms, and may even be much smaller \cite{pir19,yee13}.

It should be noted again that our assignment of Sm- or B-terminated surfaces
depends largely on the exact cleave, i.e. whether inter- or intra-octahedral
bonds are broken. Albeit the former is, as mentioned above, energetically
favorable, the latter may also occur as suggested by the observation of
so-called donuts \cite{ruan14,roe16}.

In Fig.\ \ref{defects} we present a topography over an area of 40 nm$\,
\times\,$40 nm. While we have certainly encountered areas showing a smaller
number of defects \cite{jiao16} it provides an overview of the different
types of defects found on a B-terminated surface. The largest protrusions,
\#1 (red) and \#7 (dark blue) in the upper right panel, with heights well
beyond 100 pm are most likely caused by adatoms on top of the surface. The
short red dashes in the height scan mark distances of $a$ suggesting that
the underlying lattice is not disturbed beyond the defects. Other defects,
\#5 (orange) and \#6 (black) in the lower right panel, appear to be
incorporated into the lattice as also the immediate lattice sites seem
influenced. Albeit conceivable, there is no evidence for an exchange of B by
Al in pure SmB$_6$ \cite{kon82} (note that this refers to substitution of
individual B atoms by Al, not to Al inclusion of non-negligible size
\cite{phe16,tho19}). In the Th-Pd-B system it was found that Pd may replace
two adjacent B atoms belonging to neighbouring octahedra \cite{zan94}. Along
the same line one may speculate that a similar replacement of adjacent
B atoms by impurities near the surface may result in the observed slight
displacement of surface atoms. Qualitatively different are the defects \#3
(light blue) and \#4 (magenta). Here, the lateral position (again, the
vertical dashes indicate distances of $a$) and the height oscillation of the
protrusions appear to remain unchanged while the height level is either raised
(\#3) or lowered (\#4) by about 15--20 pm over distances of about 2 lattice
constants from the center of the defect. We speculate that the defect itself
is located in a subsurface layer, possibly on a Sm site, leaving the
B-octahedra intact. It should be noted that this type of defect seems
qualitatively different from the background inhomogeneity of Fig.\
\ref{topo-def}. Albeit a clear assignment of either one of these features to
structural or electronic inhomogeneities is speculative at present, it is
obvious that a clean surface is a prerequisite for their observation. It
should also be noted that dents of about 80 pm have so far only been observed
on Sm-terminated surfaces \cite{roe14}. The topography of such dents is very
similar to the surface structure of La-terminated LaB$_6$ where La atoms are
missing from the topmost layer \cite{ozc92}. Therefore, it should be highly
instructive to investigate Sm-deficient samples Sm$_{1-x}$B$_6$ and attempt
to correlate the Sm-deficiency $x$ with the occurrence of these dents.

\subsection{EuB$_6$}
In contrast to SmB$_6$, the ferromagnetic semimetal EuB$_6$ has so far only
scarcely been investigated by STM \cite{poh18} even though its electronic
structure is not fully understood, see \cite{zha08,mas97} and references
therein. Hence, STS---in particular by using a spin-polarized tip---may
provide fresh insight. In the following, we focus on the surface topography.

\begin{figure}[t]%
\centering
\includegraphics[width=.48\textwidth]{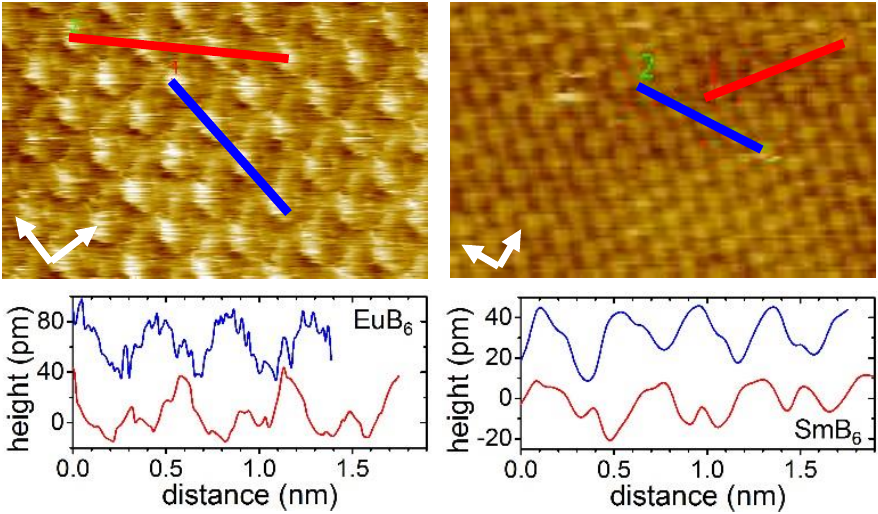}%
\caption[]{%
  Topography obtained on a Eu-terminated area (3.4 nm$\,\times\,$2.2 nm)
  of EuB$_6$ (left, $V = +0.2$ V, $I_{\rm sp} = 0.5$ nA) and a
  Sm-terminated area (5.3 nm$\,\times\,$3.4 nm) of SmB$_6$ (right,
  $V = +0.2$ V, $I_{\rm sp} = 0.6$ nA). Height scans along the
  $\langle$100$\rangle$ crystallographic directions are shown blue, and
  such parallel to the $\langle$110$\rangle$ directions
  in red. The white arrows indicate the main crystallographic directions
  $\langle 100 \rangle$ and $\langle 010 \rangle$.} \label{compEuSm}
  \end{figure}

In Fig.\ \ref{compEuSm} we compare the topographies of rare-earth terminated
samples EuB$_6$ and SmB$_6$. In both cases, atomically flat and clean surface
areas could be found after cleaving. The blue lines in the topographies
indicate where the height scans parallel to the crystallographic $\langle 100
\rangle$ directions were taken. The corrugations of heights 30--40 pm are
spaced apart by one respective lattice constant $a$. However, at the center
of the square arrangements of these main corrugations in the topography
(resulting from the cubic structure) additional humps are seen, also forming a
regular, square arrangement. This is evidenced by the red height scans along
the diagonal $\langle 110 \rangle$ directions, with the distances between the
main and the interjacent smaller corrugations corresponding to $a / \sqrt{2}$.
Based on the distances and orientations, the higher protrusions were assigned
\cite{ruan14,roe14} to the rare-earth atoms and the smaller ones to the apex
of the B octahedra, again assuming breaking inter-octahedral bonds upon
cleaving. We emphasize that the observation of interjacent smaller
corrugations along $\langle 110 \rangle$ is pivotal for the assignment of the
surface termination, yet requires sufficiently large, atomically flat and
clean surface areas. However, the consistent observation of this type of
\begin{figure}[b]%
\centering
\includegraphics*[width=0.4\textwidth]{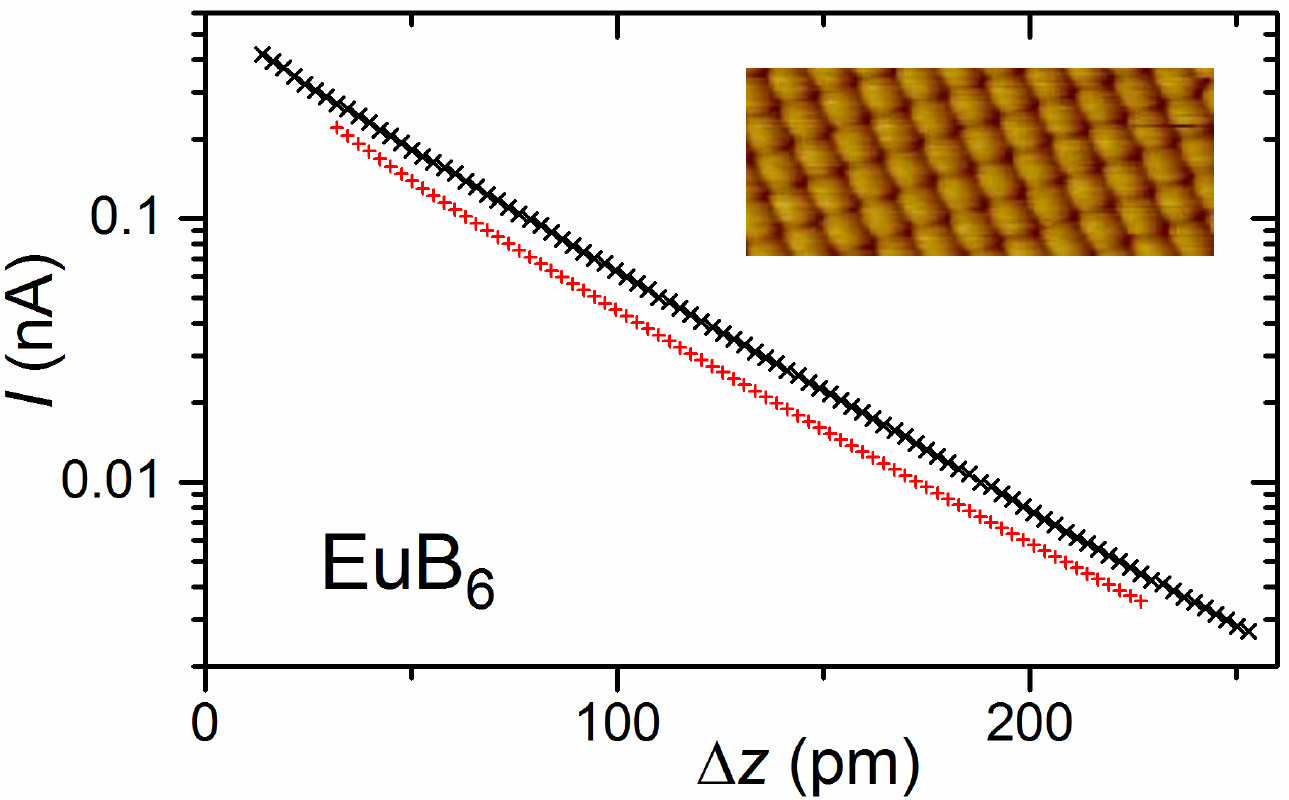}
\caption{Dependence of tunneling current $I$ on change in tip-sample distance
$\Delta z$ taken on the B-terminated surface of EuB$_6$ shown in the inset
(area 5 nm$\,\times\,$2 nm). The barrier heights correspond to 4.7 eV (black)
and 5.6 eV (red).} \label{workf}
\end{figure}
surface topography on two different members of the hexaboride family makes
a plausible case.

Based on DFT calculations it was suggested that the work function for a
Sm-terminated surface of SmB$_6$ is about 2 eV, and at least twice as high on
a B-terminated surface \cite{sun18}. We therefore started to investigate the
tunneling barrier height $\Phi$, which is related to the work functions of
\begin{figure}[t]%
\centering
\includegraphics*[width=0.48\textwidth]{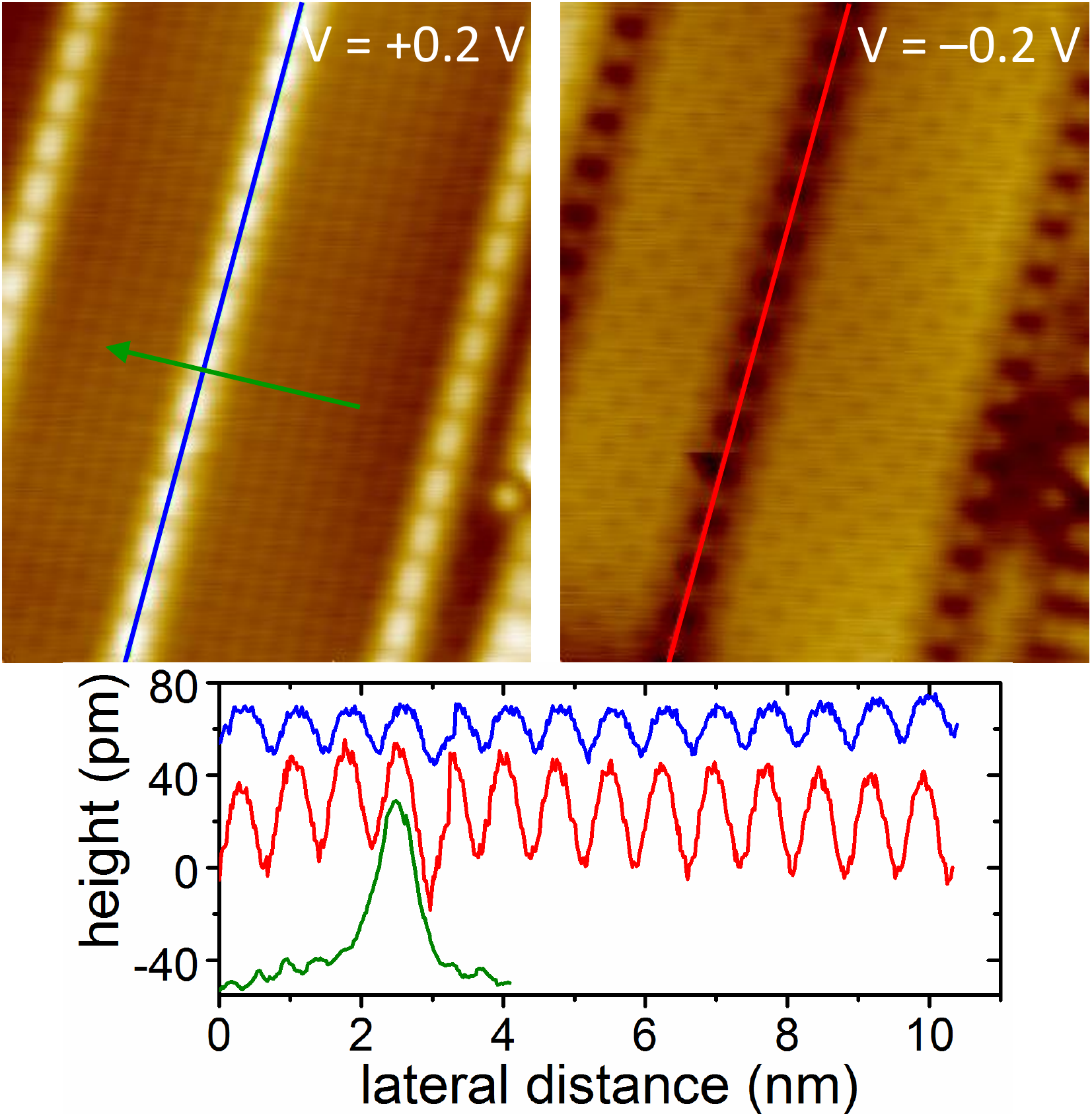}
\caption{Topography of an area of 8 nm$\,\times\,$10 nm obtained in dual-bias
mode on EuB$_6$ with Cr-tip. The line scans indicate a possible dimerization.}
\label{dimer}
\end{figure}
the sample and the tip ($\Phi_s$ and $\Phi_t$, respectively). The tunneling
current $I$ decreases exponentially with increasing tip-sample distance
$\Delta z$, i.e. $I(z) \propto\exp (-2 \kappa\, \Delta z)$. The barrier height
$\Phi$ can be calculated from $\kappa^2 = \frac{2 m_e}{\hbar^2} \Phi$, where
$m_e$ is the bare electron mass. Figure \ref{workf} shows two curves
$I(\Delta z)$ obtained on a clean B-terminated EuB$_6$ surface shown in the
inset. The barrier heights for the two exemplary curves are $\Phi =$ 4.7 eV
and 5.6 eV, i.e. they vary by almost 1 eV. Unfortunately, because of their
highly infrequent occurrence we were not able so far to measure $\Phi$ on a
Eu-terminated surface. It therefore remains to be seen whether a measurement
of the barrier height can help in identifying the termination of clean EuB$_6$
surfaces.

In the case of SmB$_6$, both the investigation of slightly Gd-substituted
samples with W tunneling tips and of pristine SmB$_6$ with Cr tips resulted in
a strong suppression of the surface state \cite{jiao18}. In fact, the d$I$/d$V$
curves in close proximity to magnetic defects and taken with magnetic tip are
akin to spectra obtained with W tip on pristine SmB$_6$ at 20~K, a temperature
high enough such that the surface states do not significantly contribute to
the tunneling spectra. These observations are expected for topologically
nontrivial surface states close to atoms carrying a sizable magnetic moment
arising from an exchange interaction \cite{liu09,wang10}. Given this
achievement in utilizing Cr tips as well as the magnetic properties of EuB$_6$
we also started to investigate surfaces of EuB$_6$ with magnetic Cr tunneling
tips. One particularly intriguing example, attained in dual-bias mode for $V
= \pm 0.2$~V, is presented in Fig.\ \ref{dimer}. The dual-bias mode is
important as an only partial contrast reversal for the two different
$V$-values is observed, rendering a position adjustment of subsequently
obtained images based solely on defects less reliable. This partial contrast
reversal also complicates the assignment of the observed features: While the
prominent bright lines seen for $V = +0.2$~V correspond in height (see green
scan line and green height profile) to Sm atoms, Fig.\ \ref{recon}, and might
be interpreted as Eu atoms, the same lines appear dark, i.e. as dents, for
$V = -0.2$~V. We note that by utilizing a magnetic tip, contrast changes may
be expected mostly on surfaces of magnetic materials. Importantly, the height
profiles measured along these lines did not show any contrast reversal upon
reversing $V$ (compare the blue and red height scans in Fig.\ \ref{dimer}).
The corrugations along these lines exhibit a periodicity of $2a$. Taken
together, one may speculate about the formation of magnetic Eu dimers on the
surface of EuB$_6$. Clear of these lines, there is no obvious
indication for a formation of such dimers: The green height scan exhibits
corrugations (away from the aforementioned line) with distances corresponding
to $a$. However, the apparent changes in contrast in areas between the line
features upon reversal of $V$ render this picture incomplete, at least.
Clearly, measurements in magnetic fields are called for, but so far we were
not able to locate such a topographic feature in our STM system with magnetic
field capabilities.

\section{Conclusions}
Investigating topographies on a large number of SmB$_6$ and EuB$_6$ samples
revealed different surface terminations which show similarities between these
two hexaborides. Such similarities are obvious for the rare-earth terminated
surfaces, a termination that is rather rare \cite{ruan14,roe14} but essential
when attempting an assignment of the different terminations. In addition,
utilizing a dual-bias mode allowed a comparison of topographies obtained with
different bias voltages on exactly identical surface areas without relying on
defects. Along with the observations of step heights less than $a$, these
observations made a reliable assignment of the rare-earth and B-terminated
surfaces possible. Apart from these atomically flat terminations, we observed
different line structures which may correspond to lines of rare-earth atoms on
top of an otherwise B-terminated surfaces. Some of these structures exhibited
intriguing properties, also if probed by magnetic tips, which warrants further
study.

\section{Acknowledgement}
We thank Silvia Seiro, Ulrich K. R\"{o}{\ss}ler, Frank Steglich, Hao Tjeng and
Jens Wiebe for support and discussions. Financial support from the Deutsche
Forschungsgemeinschaft within the priority program SPP1666 is gratefully
acknowledged. Work at Los Alamos was performed under the auspices of the U.S.
Department of Energy, Office of Basic Energy Sciences, Division of Materials
Science and Engineering.

%\bibliographystyle{psstest}
%
%
% Replace the following example bibliography with your references
% before submission:


\begin{thebibliography}{63}%
\makeatletter
\providecommand \@ifxundefined [1]{%
 \@ifx{#1\undefined}
}%
\providecommand \@ifnum [1]{%
 \ifnum #1\expandafter \@firstoftwo
 \else \expandafter \@secondoftwo
 \fi
}%
\providecommand \@ifx [1]{%
 \ifx #1\expandafter \@firstoftwo
 \else \expandafter \@secondoftwo
 \fi
}%
\providecommand \natexlab [1]{#1}%
\providecommand \enquote  [1]{``#1''}%
\providecommand \bibnamefont  [1]{#1}%
\providecommand \bibfnamefont [1]{#1}%
\providecommand \citenamefont [1]{#1}%
\providecommand \href@noop [0]{\@secondoftwo}%
\providecommand \href [0]{\begingroup \@sanitize@url \@href}%
\providecommand \@href[1]{\@@startlink{#1}\@@href}%
\providecommand \@@href[1]{\endgroup#1\@@endlink}%
\providecommand \@sanitize@url [0]{\catcode `\\12\catcode `\$12\catcode
  `\&12\catcode `\#12\catcode `\^12\catcode `\_12\catcode `\%12\relax}%
\providecommand \@@startlink[1]{}%
\providecommand \@@endlink[0]{}%
\providecommand \url  [0]{\begingroup\@sanitize@url \@url }%
\providecommand \@url [1]{\endgroup\@href {#1}{\urlprefix }}%
\providecommand \urlprefix  [0]{URL }%
\providecommand \Eprint [0]{\href }%
\providecommand \doibase [0]{http://dx.doi.org/}%
\providecommand \selectlanguage [0]{\@gobble}%
\providecommand \bibinfo  [0]{\@secondoftwo}%
\providecommand \bibfield  [0]{\@secondoftwo}%
\providecommand \translation [1]{[#1]}%
\providecommand \BibitemOpen [0]{}%
\providecommand \bibitemStop [0]{}%
\providecommand \bibitemNoStop [0]{.\EOS\space}%
\providecommand \EOS [0]{\spacefactor3000\relax}%
\providecommand \BibitemShut  [1]{\csname bibitem#1\endcsname}%
\let\auto@bib@innerbib\@empty
%</preamble>
\bibitem [{\citenamefont {Etourneau}\ and\ \citenamefont
  {Hagenmuller}(1985)}]{eto85}%
  \BibitemOpen
  \bibfield  {author} {\bibinfo {author} {\bibfnamefont {J.}~\bibnamefont
  {Etourneau}}\ and\ \bibinfo {author} {\bibfnamefont {P.}~\bibnamefont
  {Hagenmuller}},\ }\href@noop {} {\bibfield  {journal} {\bibinfo  {journal}
  {Philos. Mag.}\ }\textbf {\bibinfo {volume} {52}},\ \bibinfo {pages} {589}
  (\bibinfo {year} {1985})}\BibitemShut {NoStop}%
\bibitem [{\citenamefont {Berrada}\ \emph {et~al.}(1978)\citenamefont
  {Berrada}, \citenamefont {Mercurio}, \citenamefont {Etourneau},\ and\
  \citenamefont {Hagenmuller}}]{ber78}%
  \BibitemOpen
  \bibfield  {author} {\bibinfo {author} {\bibfnamefont {A.}~\bibnamefont
  {Berrada}}, \bibinfo {author} {\bibfnamefont {J.~P.}\ \bibnamefont
  {Mercurio}}, \bibinfo {author} {\bibfnamefont {J.}~\bibnamefont {Etourneau}},
  \ and\ \bibinfo {author} {\bibfnamefont {P.}~\bibnamefont {Hagenmuller}},\
  }\href@noop {} {\bibfield  {journal} {\bibinfo  {journal} {J. Less-Common
  Met.}\ }\textbf {\bibinfo {volume} {59}},\ \bibinfo {pages} {7} (\bibinfo
  {year} {1978})}\BibitemShut {NoStop}%
\bibitem [{\citenamefont {Young}\ \emph {et~al.}(1999)\citenamefont {Young},
  \citenamefont {Hall}, \citenamefont {Torelli}, \citenamefont {Fisk},
  \citenamefont {Sarrao}, \citenamefont {Thompson}, \citenamefont {Ott},
  \citenamefont {Oseroff}, \citenamefont {Goodrich},\ and\ \citenamefont
  {Zysler}}]{you99}%
  \BibitemOpen
  \bibfield  {author} {\bibinfo {author} {\bibfnamefont {D.~P.}\ \bibnamefont
  {Young}}, \bibinfo {author} {\bibfnamefont {D.}~\bibnamefont {Hall}},
  \bibinfo {author} {\bibfnamefont {M.~E.}\ \bibnamefont {Torelli}}, \bibinfo
  {author} {\bibfnamefont {Z.}~\bibnamefont {Fisk}}, \bibinfo {author}
  {\bibfnamefont {J.~L.}\ \bibnamefont {Sarrao}}, \bibinfo {author}
  {\bibfnamefont {J.~D.}\ \bibnamefont {Thompson}}, \bibinfo {author}
  {\bibfnamefont {H.~R.}\ \bibnamefont {Ott}}, \bibinfo {author} {\bibfnamefont
  {S.~B.}\ \bibnamefont {Oseroff}}, \bibinfo {author} {\bibfnamefont {R.~G.}\
  \bibnamefont {Goodrich}}, \ and\ \bibinfo {author} {\bibfnamefont
  {R.}~\bibnamefont {Zysler}},\ }\href@noop {} {\bibfield  {journal} {\bibinfo
  {journal} {Nature}\ }\textbf {\bibinfo {volume} {397}},\ \bibinfo {pages}
  {412} (\bibinfo {year} {1999})}\BibitemShut {NoStop}%
\bibitem [{\citenamefont {Stankiewicz}\ \emph {et~al.}(2014)\citenamefont
  {Stankiewicz}, \citenamefont {Ses{\'e}}, \citenamefont {Balakrishnan},\ and\
  \citenamefont {Fisk}}]{sta14}%
  \BibitemOpen
  \bibfield  {author} {\bibinfo {author} {\bibfnamefont {J.}~\bibnamefont
  {Stankiewicz}}, \bibinfo {author} {\bibfnamefont {J.}~\bibnamefont
  {Ses{\'e}}}, \bibinfo {author} {\bibfnamefont {G.}~\bibnamefont
  {Balakrishnan}}, \ and\ \bibinfo {author} {\bibfnamefont {Z.}~\bibnamefont
  {Fisk}},\ }\href@noop {} {\bibfield  {journal} {\bibinfo  {journal} {Phys.\
  Rev.\ B}\ }\textbf {\bibinfo {volume} {90}},\ \bibinfo {pages} {155128}
  (\bibinfo {year} {2014})}\BibitemShut {NoStop}%
\bibitem [{\citenamefont {Effantin}\ \emph {et~al.}(1985)\citenamefont
  {Effantin}, \citenamefont {Rossatmignod}, \citenamefont {Burlet},
  \citenamefont {Bartholin}, \citenamefont {Kunii},\ and\ \citenamefont
  {Kasuya}}]{eff85}%
  \BibitemOpen
  \bibfield  {author} {\bibinfo {author} {\bibfnamefont {J.~M.}\ \bibnamefont
  {Effantin}}, \bibinfo {author} {\bibfnamefont {J.}~\bibnamefont
  {Rossatmignod}}, \bibinfo {author} {\bibfnamefont {P.}~\bibnamefont
  {Burlet}}, \bibinfo {author} {\bibfnamefont {H.}~\bibnamefont {Bartholin}},
  \bibinfo {author} {\bibfnamefont {S.}~\bibnamefont {Kunii}}, \ and\ \bibinfo
  {author} {\bibfnamefont {T.}~\bibnamefont {Kasuya}},\ }\href@noop {}
  {\bibfield  {journal} {\bibinfo  {journal} {J. Magn. Magn. Mater.}\ }\textbf
  {\bibinfo {volume} {47-48}},\ \bibinfo {pages} {145} (\bibinfo {year}
  {1985})}\BibitemShut {NoStop}%
\bibitem [{\citenamefont {Grushko}\ \emph {et~al.}(1985)\citenamefont
  {Grushko}, \citenamefont {Paderno}, \citenamefont {Mishin}, \citenamefont
  {Molkanov}, \citenamefont {Shadrina}, \citenamefont {Konovalova},\ and\
  \citenamefont {Dudnik}}]{gru85}%
  \BibitemOpen
  \bibfield  {author} {\bibinfo {author} {\bibfnamefont {Y.~S.}\ \bibnamefont
  {Grushko}}, \bibinfo {author} {\bibfnamefont {Y.~B.}\ \bibnamefont
  {Paderno}}, \bibinfo {author} {\bibfnamefont {K.~Y.}\ \bibnamefont {Mishin}},
  \bibinfo {author} {\bibfnamefont {L.~I.}\ \bibnamefont {Molkanov}}, \bibinfo
  {author} {\bibfnamefont {G.~A.}\ \bibnamefont {Shadrina}}, \bibinfo {author}
  {\bibfnamefont {E.~S.}\ \bibnamefont {Konovalova}}, \ and\ \bibinfo {author}
  {\bibfnamefont {E.~M.}\ \bibnamefont {Dudnik}},\ }\href@noop {} {\bibfield
  {journal} {\bibinfo  {journal} {Phys. Stat. Sol. B}\ }\textbf {\bibinfo
  {volume} {128}},\ \bibinfo {pages} {591} (\bibinfo {year}
  {1985})}\BibitemShut {NoStop}%
\bibitem [{\citenamefont {Vainshtein}\ \emph {et~al.}(1965)\citenamefont
  {Vainshtein}, \citenamefont {Blokhin},\ and\ \citenamefont
  {Paderno}}]{vai64}%
  \BibitemOpen
  \bibfield  {author} {\bibinfo {author} {\bibfnamefont {E.~E.}\ \bibnamefont
  {Vainshtein}}, \bibinfo {author} {\bibfnamefont {S.~M.}\ \bibnamefont
  {Blokhin}}, \ and\ \bibinfo {author} {\bibfnamefont {Y.~B.}\ \bibnamefont
  {Paderno}},\ }\href@noop {} {\bibfield  {journal} {\bibinfo  {journal} {Sov.\
  Phys.-Solid State}\ }\textbf {\bibinfo {volume} {6}},\ \bibinfo {pages}
  {2318} (\bibinfo {year} {1965})}\BibitemShut {NoStop}%
\bibitem [{\citenamefont {Utsumi}\ \emph {et~al.}(2017)\citenamefont {Utsumi},
  \citenamefont {Kasinathan}, \citenamefont {Ko}, \citenamefont {Agrestini},
  \citenamefont {Haverkort}, \citenamefont {Wirth}, \citenamefont {Wu},
  \citenamefont {Tsuei}, \citenamefont {Kim}, \citenamefont {Fisk},
  \citenamefont {Tanaka}, \citenamefont {Thalmeier},\ and\ \citenamefont
  {Tjeng}}]{uts17}%
  \BibitemOpen
  \bibfield  {author} {\bibinfo {author} {\bibfnamefont {Y.}~\bibnamefont
  {Utsumi}}, \bibinfo {author} {\bibfnamefont {D.}~\bibnamefont {Kasinathan}},
  \bibinfo {author} {\bibfnamefont {K.-T.}\ \bibnamefont {Ko}}, \bibinfo
  {author} {\bibfnamefont {S.}~\bibnamefont {Agrestini}}, \bibinfo {author}
  {\bibfnamefont {M.~W.}\ \bibnamefont {Haverkort}}, \bibinfo {author}
  {\bibfnamefont {S.}~\bibnamefont {Wirth}}, \bibinfo {author} {\bibfnamefont
  {Y.-H.}\ \bibnamefont {Wu}}, \bibinfo {author} {\bibfnamefont {K.-D.}\
  \bibnamefont {Tsuei}}, \bibinfo {author} {\bibfnamefont {D.-J.}\ \bibnamefont
  {Kim}}, \bibinfo {author} {\bibfnamefont {Z.}~\bibnamefont {Fisk}}, \bibinfo
  {author} {\bibfnamefont {A.}~\bibnamefont {Tanaka}}, \bibinfo {author}
  {\bibfnamefont {P.}~\bibnamefont {Thalmeier}}, \ and\ \bibinfo {author}
  {\bibfnamefont {L.~H.}\ \bibnamefont {Tjeng}},\ }\href@noop {} {\bibfield
  {journal} {\bibinfo  {journal} {Phys. Rev. B}\ }\textbf {\bibinfo {volume}
  {96}},\ \bibinfo {pages} {155130} (\bibinfo {year} {2017})}\BibitemShut
  {NoStop}%
\bibitem [{\citenamefont {Dzero}\ \emph {et~al.}(2010)\citenamefont {Dzero},
  \citenamefont {Sun}, \citenamefont {Galitski},\ and\ \citenamefont
  {Coleman}}]{dze10}%
  \BibitemOpen
  \bibfield  {author} {\bibinfo {author} {\bibfnamefont {M.}~\bibnamefont
  {Dzero}}, \bibinfo {author} {\bibfnamefont {K.}~\bibnamefont {Sun}}, \bibinfo
  {author} {\bibfnamefont {V.}~\bibnamefont {Galitski}}, \ and\ \bibinfo
  {author} {\bibfnamefont {P.}~\bibnamefont {Coleman}},\ }\href@noop {}
  {\bibfield  {journal} {\bibinfo  {journal} {Phys.\ Rev.\ Lett.}\ }\textbf
  {\bibinfo {volume} {104}},\ \bibinfo {pages} {106408} (\bibinfo {year}
  {2010})}\BibitemShut {NoStop}%
\bibitem [{\citenamefont {Aeppli}\ and\ \citenamefont {Fisk}(1992)}]{aep92}%
  \BibitemOpen
  \bibfield  {author} {\bibinfo {author} {\bibfnamefont {G.}~\bibnamefont
  {Aeppli}}\ and\ \bibinfo {author} {\bibfnamefont {Z.}~\bibnamefont {Fisk}},\
  }\href@noop {} {\bibfield  {journal} {\bibinfo  {journal} {Comments Cond.\
  Mat.\ Phys.}\ }\textbf {\bibinfo {volume} {16}},\ \bibinfo {pages} {155}
  (\bibinfo {year} {1992})}\BibitemShut {NoStop}%
\bibitem [{\citenamefont {Riseborough}(2000)}]{ris00}%
  \BibitemOpen
  \bibfield  {author} {\bibinfo {author} {\bibfnamefont {P.~S.}\ \bibnamefont
  {Riseborough}},\ }\href@noop {} {\bibfield  {journal} {\bibinfo  {journal}
  {Adv.\ Phys.}\ }\textbf {\bibinfo {volume} {49}},\ \bibinfo {pages} {257}
  (\bibinfo {year} {2000})}\BibitemShut {NoStop}%
\bibitem [{\citenamefont {Takimoto}(2011)}]{tak11}%
  \BibitemOpen
  \bibfield  {author} {\bibinfo {author} {\bibfnamefont {T.}~\bibnamefont
  {Takimoto}},\ }\href@noop {} {\bibfield  {journal} {\bibinfo  {journal} {J.\
  Phys.\ Soc.\ Jpn.}\ }\textbf {\bibinfo {volume} {80}},\ \bibinfo {pages}
  {123710} (\bibinfo {year} {2011})}\BibitemShut {NoStop}%
\bibitem [{\citenamefont {Lu}\ \emph {et~al.}(2013)\citenamefont {Lu},
  \citenamefont {Zhao}, \citenamefont {Weng}, \citenamefont {Fang},\ and\
  \citenamefont {Dai}}]{lu13}%
  \BibitemOpen
  \bibfield  {author} {\bibinfo {author} {\bibfnamefont {F.}~\bibnamefont
  {Lu}}, \bibinfo {author} {\bibfnamefont {J.}~\bibnamefont {Zhao}}, \bibinfo
  {author} {\bibfnamefont {H.}~\bibnamefont {Weng}}, \bibinfo {author}
  {\bibfnamefont {Z.}~\bibnamefont {Fang}}, \ and\ \bibinfo {author}
  {\bibfnamefont {X.}~\bibnamefont {Dai}},\ }\href@noop {} {\bibfield
  {journal} {\bibinfo  {journal} {Phys.\ Rev.\ Lett.}\ }\textbf {\bibinfo
  {volume} {110}},\ \bibinfo {pages} {096401} (\bibinfo {year}
  {2013})}\BibitemShut {NoStop}%
\bibitem [{\citenamefont {Kim}\ \emph {et~al.}(2014{\natexlab{a}})\citenamefont
  {Kim}, \citenamefont {Kim}, \citenamefont {Kang}, \citenamefont {Kim},
  \citenamefont {Choi}, \citenamefont {Kang}, \citenamefont {Denlinger},\ and\
  \citenamefont {Min}}]{jkim14}%
  \BibitemOpen
  \bibfield  {author} {\bibinfo {author} {\bibfnamefont {J.}~\bibnamefont
  {Kim}}, \bibinfo {author} {\bibfnamefont {K.}~\bibnamefont {Kim}}, \bibinfo
  {author} {\bibfnamefont {C.-J.}\ \bibnamefont {Kang}}, \bibinfo {author}
  {\bibfnamefont {S.}~\bibnamefont {Kim}}, \bibinfo {author} {\bibfnamefont
  {H.~C.}\ \bibnamefont {Choi}}, \bibinfo {author} {\bibfnamefont {J.-S.}\
  \bibnamefont {Kang}}, \bibinfo {author} {\bibfnamefont {J.~D.}\ \bibnamefont
  {Denlinger}}, \ and\ \bibinfo {author} {\bibfnamefont {B.~I.}\ \bibnamefont
  {Min}},\ }\href@noop {} {\bibfield  {journal} {\bibinfo  {journal} {Phys.
  Rev. B}\ }\textbf {\bibinfo {volume} {90}},\ \bibinfo {pages} {075131}
  (\bibinfo {year} {2014}{\natexlab{a}})}\BibitemShut {NoStop}%
\bibitem [{\citenamefont {Xu}\ \emph {et~al.}(2014)\citenamefont {Xu},
  \citenamefont {Biswas}, \citenamefont {Dil}, \citenamefont {Dhaka},
  \citenamefont {Landolt}, \citenamefont {Muff}, \citenamefont {Matt},
  \citenamefont {Shi}, \citenamefont {Plumb}, \citenamefont {Radovi{\'c}},
  \citenamefont {Pomjakushina}, \citenamefont {Conder}, \citenamefont {Amato},
  \citenamefont {Borisenko}, \citenamefont {Yu}, \citenamefont {Weng},
  \citenamefont {Fang}, \citenamefont {Dai}, \citenamefont {Mesot},
  \citenamefont {Ding},\ and\ \citenamefont {Shi}}]{nxu14}%
  \BibitemOpen
  \bibfield  {author} {\bibinfo {author} {\bibfnamefont {N.}~\bibnamefont
  {Xu}}, \bibinfo {author} {\bibfnamefont {P.~K.}\ \bibnamefont {Biswas}},
  \bibinfo {author} {\bibfnamefont {J.~H.}\ \bibnamefont {Dil}}, \bibinfo
  {author} {\bibfnamefont {R.~S.}\ \bibnamefont {Dhaka}}, \bibinfo {author}
  {\bibfnamefont {G.}~\bibnamefont {Landolt}}, \bibinfo {author} {\bibfnamefont
  {S.}~\bibnamefont {Muff}}, \bibinfo {author} {\bibfnamefont {C.~E.}\
  \bibnamefont {Matt}}, \bibinfo {author} {\bibfnamefont {X.}~\bibnamefont
  {Shi}}, \bibinfo {author} {\bibfnamefont {N.~C.}\ \bibnamefont {Plumb}},
  \bibinfo {author} {\bibfnamefont {M.}~\bibnamefont {Radovi{\'c}}}, \bibinfo
  {author} {\bibfnamefont {E.}~\bibnamefont {Pomjakushina}}, \bibinfo {author}
  {\bibfnamefont {K.}~\bibnamefont {Conder}}, \bibinfo {author} {\bibfnamefont
  {A.}~\bibnamefont {Amato}}, \bibinfo {author} {\bibfnamefont {S.~V.}\
  \bibnamefont {Borisenko}}, \bibinfo {author} {\bibfnamefont {R.}~\bibnamefont
  {Yu}}, \bibinfo {author} {\bibfnamefont {H.~M.}\ \bibnamefont {Weng}},
  \bibinfo {author} {\bibfnamefont {Z.}~\bibnamefont {Fang}}, \bibinfo {author}
  {\bibfnamefont {X.}~\bibnamefont {Dai}}, \bibinfo {author} {\bibfnamefont
  {J.}~\bibnamefont {Mesot}}, \bibinfo {author} {\bibfnamefont
  {H.}~\bibnamefont {Ding}}, \ and\ \bibinfo {author} {\bibfnamefont
  {M.}~\bibnamefont {Shi}},\ }\href@noop {} {\bibfield  {journal} {\bibinfo
  {journal} {Nature Commun.}\ }\textbf {\bibinfo {volume} {5}},\ \bibinfo
  {pages} {4566} (\bibinfo {year} {2014})}\BibitemShut {NoStop}%
\bibitem [{\citenamefont {Suga}\ \emph {et~al.}(2014)\citenamefont {Suga},
  \citenamefont {Sakamoto}, \citenamefont {Okuda}, \citenamefont {Miyamoto},
  \citenamefont {Kuroda}, \citenamefont {Sekiyama}, \citenamefont {Yamaguchi},
  \citenamefont {Fujiwara}, \citenamefont {Irizawa}, \citenamefont {Ito},
  \citenamefont {Kimura}, \citenamefont {Balashov}, \citenamefont {Wulfhekel},
  \citenamefont {Yeo}, \citenamefont {Iga},\ and\ \citenamefont
  {Imada}}]{sug14}%
  \BibitemOpen
  \bibfield  {author} {\bibinfo {author} {\bibfnamefont {S.}~\bibnamefont
  {Suga}}, \bibinfo {author} {\bibfnamefont {K.}~\bibnamefont {Sakamoto}},
  \bibinfo {author} {\bibfnamefont {T.}~\bibnamefont {Okuda}}, \bibinfo
  {author} {\bibfnamefont {K.}~\bibnamefont {Miyamoto}}, \bibinfo {author}
  {\bibfnamefont {K.}~\bibnamefont {Kuroda}}, \bibinfo {author} {\bibfnamefont
  {A.}~\bibnamefont {Sekiyama}}, \bibinfo {author} {\bibfnamefont
  {J.}~\bibnamefont {Yamaguchi}}, \bibinfo {author} {\bibfnamefont
  {H.}~\bibnamefont {Fujiwara}}, \bibinfo {author} {\bibfnamefont
  {A.}~\bibnamefont {Irizawa}}, \bibinfo {author} {\bibfnamefont
  {T.}~\bibnamefont {Ito}}, \bibinfo {author} {\bibfnamefont {S.}~\bibnamefont
  {Kimura}}, \bibinfo {author} {\bibfnamefont {T.}~\bibnamefont {Balashov}},
  \bibinfo {author} {\bibfnamefont {W.}~\bibnamefont {Wulfhekel}}, \bibinfo
  {author} {\bibfnamefont {S.}~\bibnamefont {Yeo}}, \bibinfo {author}
  {\bibfnamefont {F.}~\bibnamefont {Iga}}, \ and\ \bibinfo {author}
  {\bibfnamefont {S.}~\bibnamefont {Imada}},\ }\href@noop {} {\bibfield
  {journal} {\bibinfo  {journal} {J.\ Phys.\ Soc.\ Jpn.}\ }\textbf {\bibinfo
  {volume} {83}},\ \bibinfo {pages} {014705} (\bibinfo {year}
  {2014})}\BibitemShut {NoStop}%
\bibitem [{\citenamefont {Wolgast}\ \emph {et~al.}(2013)\citenamefont
  {Wolgast}, \citenamefont {Kurdak}, \citenamefont {Sun}, \citenamefont
  {Allen}, \citenamefont {Kim},\ and\ \citenamefont {Fisk}}]{wol13}%
  \BibitemOpen
  \bibfield  {author} {\bibinfo {author} {\bibfnamefont {S.}~\bibnamefont
  {Wolgast}}, \bibinfo {author} {\bibfnamefont {C.}~\bibnamefont {Kurdak}},
  \bibinfo {author} {\bibfnamefont {K.}~\bibnamefont {Sun}}, \bibinfo {author}
  {\bibfnamefont {J.~W.}\ \bibnamefont {Allen}}, \bibinfo {author}
  {\bibfnamefont {D.-J.}\ \bibnamefont {Kim}}, \ and\ \bibinfo {author}
  {\bibfnamefont {Z.}~\bibnamefont {Fisk}},\ }\href@noop {} {\bibfield
  {journal} {\bibinfo  {journal} {Phys.\ Rev.\ B}\ }\textbf {\bibinfo {volume}
  {88}},\ \bibinfo {pages} {180405(R)} (\bibinfo {year} {2013})}\BibitemShut
  {NoStop}%
\bibitem [{\citenamefont {Eo}\ \emph {et~al.}(2019)\citenamefont {Eo},
  \citenamefont {Rakoski}, \citenamefont {Lucien}, \citenamefont {Mihaliov},
  \citenamefont {Kurdak}, \citenamefont {Rosa},\ and\ \citenamefont
  {Fisk}}]{eo19}%
  \BibitemOpen
  \bibfield  {author} {\bibinfo {author} {\bibfnamefont {Y.~S.}\ \bibnamefont
  {Eo}}, \bibinfo {author} {\bibfnamefont {A.}~\bibnamefont {Rakoski}},
  \bibinfo {author} {\bibfnamefont {J.}~\bibnamefont {Lucien}}, \bibinfo
  {author} {\bibfnamefont {D.}~\bibnamefont {Mihaliov}}, \bibinfo {author}
  {\bibfnamefont {C.}~\bibnamefont {Kurdak}}, \bibinfo {author} {\bibfnamefont
  {P.~F.~S.}\ \bibnamefont {Rosa}}, \ and\ \bibinfo {author} {\bibfnamefont
  {Z.}~\bibnamefont {Fisk}},\ }\href@noop {} {\bibfield  {journal} {\bibinfo
  {journal} {Proc.\ Natl.\ Acad.\ Sci.\ USA}\ }\textbf {\bibinfo {volume}
  {116}},\ \bibinfo {pages} {12638} (\bibinfo {year} {2019})}\BibitemShut
  {NoStop}%
\bibitem [{\citenamefont {Hlawenka}\ \emph {et~al.}(2018)\citenamefont
  {Hlawenka}, \citenamefont {Siemensmeyer}, \citenamefont {Weschke},
  \citenamefont {Varykhalov}, \citenamefont {S\'{a}nchez-Barriga},
  \citenamefont {Shitsevalova}, \citenamefont {Dukhnenko}, \citenamefont
  {Filipov}, \citenamefont {Gab\'{a}ni}, \citenamefont {Flachbart},
  \citenamefont {Rader},\ and\ \citenamefont {Rienks}}]{hla18}%
  \BibitemOpen
  \bibfield  {author} {\bibinfo {author} {\bibfnamefont {P.}~\bibnamefont
  {Hlawenka}}, \bibinfo {author} {\bibfnamefont {K.}~\bibnamefont
  {Siemensmeyer}}, \bibinfo {author} {\bibfnamefont {E.}~\bibnamefont
  {Weschke}}, \bibinfo {author} {\bibfnamefont {A.}~\bibnamefont {Varykhalov}},
  \bibinfo {author} {\bibfnamefont {J.}~\bibnamefont {S\'{a}nchez-Barriga}},
  \bibinfo {author} {\bibfnamefont {N.~Y.}\ \bibnamefont {Shitsevalova}},
  \bibinfo {author} {\bibfnamefont {A.~V.}\ \bibnamefont {Dukhnenko}}, \bibinfo
  {author} {\bibfnamefont {V.~B.}\ \bibnamefont {Filipov}}, \bibinfo {author}
  {\bibfnamefont {S.}~\bibnamefont {Gab\'{a}ni}}, \bibinfo {author}
  {\bibfnamefont {K.}~\bibnamefont {Flachbart}}, \bibinfo {author}
  {\bibfnamefont {O.}~\bibnamefont {Rader}}, \ and\ \bibinfo {author}
  {\bibfnamefont {E.~D.~L.}\ \bibnamefont {Rienks}},\ }\href@noop {} {\bibfield
   {journal} {\bibinfo  {journal} {Nature Commun.}\ }\textbf {\bibinfo {volume}
  {9}},\ \bibinfo {pages} {517} (\bibinfo {year} {2018})}\BibitemShut {NoStop}%
\bibitem [{\citenamefont {Baruselli}\ and\ \citenamefont
  {Vojta}(2015)}]{bar15}%
  \BibitemOpen
  \bibfield  {author} {\bibinfo {author} {\bibfnamefont {P.~P.}\ \bibnamefont
  {Baruselli}}\ and\ \bibinfo {author} {\bibfnamefont {M.}~\bibnamefont
  {Vojta}},\ }\href@noop {} {\bibfield  {journal} {\bibinfo  {journal} {Phys.\
  Rev.\ Lett.}\ }\textbf {\bibinfo {volume} {115}},\ \bibinfo {pages} {156404}
  (\bibinfo {year} {2015})}\BibitemShut {NoStop}%
\bibitem [{\citenamefont {Legner}\ \emph {et~al.}(2015)\citenamefont {Legner},
  \citenamefont {R{\"u}egg},\ and\ \citenamefont {Sigrist}}]{leg15}%
  \BibitemOpen
  \bibfield  {author} {\bibinfo {author} {\bibfnamefont {M.}~\bibnamefont
  {Legner}}, \bibinfo {author} {\bibfnamefont {A.}~\bibnamefont {R{\"u}egg}}, \
  and\ \bibinfo {author} {\bibfnamefont {M.}~\bibnamefont {Sigrist}},\
  }\href@noop {} {\bibfield  {journal} {\bibinfo  {journal} {Phys.\ Rev.\
  Lett.}\ }\textbf {\bibinfo {volume} {115}},\ \bibinfo {pages} {156405}
  (\bibinfo {year} {2015})}\BibitemShut {NoStop}%
\bibitem [{\citenamefont {Sundermann}\ \emph {et~al.}(2018)\citenamefont
  {Sundermann}, \citenamefont {Yavas}, \citenamefont {Chen}, \citenamefont
  {Kim}, \citenamefont {Fisk}, \citenamefont {Kasinathan}, \citenamefont
  {Haverkort}, \citenamefont {Thalmeier}, \citenamefont {Severing},\ and\
  \citenamefont {Tjeng}}]{sev18}%
  \BibitemOpen
  \bibfield  {author} {\bibinfo {author} {\bibfnamefont {M.}~\bibnamefont
  {Sundermann}}, \bibinfo {author} {\bibfnamefont {H.}~\bibnamefont {Yavas}},
  \bibinfo {author} {\bibfnamefont {K.}~\bibnamefont {Chen}}, \bibinfo {author}
  {\bibfnamefont {D.}~\bibnamefont {Kim}}, \bibinfo {author} {\bibfnamefont
  {Z.}~\bibnamefont {Fisk}}, \bibinfo {author} {\bibfnamefont {D.}~\bibnamefont
  {Kasinathan}}, \bibinfo {author} {\bibfnamefont {M.}~\bibnamefont
  {Haverkort}}, \bibinfo {author} {\bibfnamefont {P.}~\bibnamefont
  {Thalmeier}}, \bibinfo {author} {\bibfnamefont {A.}~\bibnamefont {Severing}},
  \ and\ \bibinfo {author} {\bibfnamefont {L.}~\bibnamefont {Tjeng}},\
  }\href@noop {} {\bibfield  {journal} {\bibinfo  {journal} {Phys.\ Rev.\
  Lett.}\ }\textbf {\bibinfo {volume} {120}},\ \bibinfo {pages} {016402}
  (\bibinfo {year} {2018})}\BibitemShut {NoStop}%
\bibitem [{\citenamefont {Antonov}\ \emph {et~al.}(2002)\citenamefont
  {Antonov}, \citenamefont {Harmon},\ and\ \citenamefont {Yaresko}}]{ant02}%
  \BibitemOpen
  \bibfield  {author} {\bibinfo {author} {\bibfnamefont {V.~N.}\ \bibnamefont
  {Antonov}}, \bibinfo {author} {\bibfnamefont {B.~N.}\ \bibnamefont {Harmon}},
  \ and\ \bibinfo {author} {\bibfnamefont {A.~N.}\ \bibnamefont {Yaresko}},\
  }\href@noop {} {\bibfield  {journal} {\bibinfo  {journal} {Phys.\ Rev.\ B}\
  }\textbf {\bibinfo {volume} {66}},\ \bibinfo {pages} {165209} (\bibinfo
  {year} {2002})}\BibitemShut {NoStop}%
\bibitem [{\citenamefont {Kang}\ \emph {et~al.}(2015)\citenamefont {Kang},
  \citenamefont {Kim}, \citenamefont {Kim}, \citenamefont {Kang}, \citenamefont
  {Denlinger},\ and\ \citenamefont {Min}}]{kan15}%
  \BibitemOpen
  \bibfield  {author} {\bibinfo {author} {\bibfnamefont {C.-J.}\ \bibnamefont
  {Kang}}, \bibinfo {author} {\bibfnamefont {J.}~\bibnamefont {Kim}}, \bibinfo
  {author} {\bibfnamefont {K.}~\bibnamefont {Kim}}, \bibinfo {author}
  {\bibfnamefont {J.}~\bibnamefont {Kang}}, \bibinfo {author} {\bibfnamefont
  {J.~D.}\ \bibnamefont {Denlinger}}, \ and\ \bibinfo {author} {\bibfnamefont
  {B.~I.}\ \bibnamefont {Min}},\ }\href@noop {} {\bibfield  {journal} {\bibinfo
   {journal} {J.\ Phys.\ Soc.\ Jpn.}\ }\textbf {\bibinfo {volume} {84}},\
  \bibinfo {pages} {024722} (\bibinfo {year} {2015})}\BibitemShut {NoStop}%
\bibitem [{\citenamefont {Min}\ \emph {et~al.}(2017)\citenamefont {Min},
  \citenamefont {Goth}, \citenamefont {Lutz}, \citenamefont {Bentmann},
  \citenamefont {Kang}, \citenamefont {Cho}, \citenamefont {Werner},
  \citenamefont {Chen}, \citenamefont {Assaad},\ and\ \citenamefont
  {Reinert}}]{min17}%
  \BibitemOpen
  \bibfield  {author} {\bibinfo {author} {\bibfnamefont {C.-H.}\ \bibnamefont
  {Min}}, \bibinfo {author} {\bibfnamefont {F.}~\bibnamefont {Goth}}, \bibinfo
  {author} {\bibfnamefont {P.}~\bibnamefont {Lutz}}, \bibinfo {author}
  {\bibfnamefont {H.}~\bibnamefont {Bentmann}}, \bibinfo {author}
  {\bibfnamefont {B.~Y.}\ \bibnamefont {Kang}}, \bibinfo {author}
  {\bibfnamefont {B.~K.}\ \bibnamefont {Cho}}, \bibinfo {author} {\bibfnamefont
  {J.}~\bibnamefont {Werner}}, \bibinfo {author} {\bibfnamefont {K.-S.}\
  \bibnamefont {Chen}}, \bibinfo {author} {\bibfnamefont {F.}~\bibnamefont
  {Assaad}}, \ and\ \bibinfo {author} {\bibfnamefont {F.}~\bibnamefont
  {Reinert}},\ }\href@noop {} {\bibfield  {journal} {\bibinfo  {journal} {Sci.
  Rep.}\ }\textbf {\bibinfo {volume} {7}},\ \bibinfo {pages} {11980} (\bibinfo
  {year} {2017})}\BibitemShut {NoStop}%
\bibitem [{\citenamefont {Zhu}\ \emph {et~al.}(2013)\citenamefont {Zhu},
  \citenamefont {Nicolaou}, \citenamefont {Levy}, \citenamefont {Butch},
  \citenamefont {Syers}, \citenamefont {Wang}, \citenamefont {Paglione},
  \citenamefont {Sawatzky}, \citenamefont {Elfimov},\ and\ \citenamefont
  {Damascelli}}]{zhu13}%
  \BibitemOpen
  \bibfield  {author} {\bibinfo {author} {\bibfnamefont {Z.-H.}\ \bibnamefont
  {Zhu}}, \bibinfo {author} {\bibfnamefont {A.}~\bibnamefont {Nicolaou}},
  \bibinfo {author} {\bibfnamefont {G.}~\bibnamefont {Levy}}, \bibinfo {author}
  {\bibfnamefont {N.~P.}\ \bibnamefont {Butch}}, \bibinfo {author}
  {\bibfnamefont {P.}~\bibnamefont {Syers}}, \bibinfo {author} {\bibfnamefont
  {X.~F.}\ \bibnamefont {Wang}}, \bibinfo {author} {\bibfnamefont
  {J.}~\bibnamefont {Paglione}}, \bibinfo {author} {\bibfnamefont {G.~A.}\
  \bibnamefont {Sawatzky}}, \bibinfo {author} {\bibfnamefont {I.~S.}\
  \bibnamefont {Elfimov}}, \ and\ \bibinfo {author} {\bibfnamefont
  {A.}~\bibnamefont {Damascelli}},\ }\href@noop {} {\bibfield  {journal}
  {\bibinfo  {journal} {Phys.\ Rev.\ Lett.}\ }\textbf {\bibinfo {volume}
  {111}},\ \bibinfo {pages} {216402} (\bibinfo {year} {2013})}\BibitemShut
  {NoStop}%
\bibitem [{\citenamefont {Ruan}\ \emph {et~al.}(2014)\citenamefont {Ruan},
  \citenamefont {Ye}, \citenamefont {Guo}, \citenamefont {Chen}, \citenamefont
  {Chen}, \citenamefont {Zhang},\ and\ \citenamefont {Wang}}]{ruan14}%
  \BibitemOpen
  \bibfield  {author} {\bibinfo {author} {\bibfnamefont {W.}~\bibnamefont
  {Ruan}}, \bibinfo {author} {\bibfnamefont {C.}~\bibnamefont {Ye}}, \bibinfo
  {author} {\bibfnamefont {M.}~\bibnamefont {Guo}}, \bibinfo {author}
  {\bibfnamefont {F.}~\bibnamefont {Chen}}, \bibinfo {author} {\bibfnamefont
  {X.}~\bibnamefont {Chen}}, \bibinfo {author} {\bibfnamefont {G.-M.}\
  \bibnamefont {Zhang}}, \ and\ \bibinfo {author} {\bibfnamefont
  {Y.}~\bibnamefont {Wang}},\ }\href@noop {} {\bibfield  {journal} {\bibinfo
  {journal} {Phys.\ Rev.\ Lett.}\ }\textbf {\bibinfo {volume} {112}},\ \bibinfo
  {pages} {136401} (\bibinfo {year} {2014})}\BibitemShut {NoStop}%
\bibitem [{\citenamefont {R{\"o}{\ss}ler}\ \emph {et~al.}(2014)\citenamefont
  {R{\"o}{\ss}ler}, \citenamefont {Jang}, \citenamefont {Kim}, \citenamefont
  {Tjeng}, \citenamefont {Fisk}, \citenamefont {Steglich},\ and\ \citenamefont
  {Wirth}}]{roe14}%
  \BibitemOpen
  \bibfield  {author} {\bibinfo {author} {\bibfnamefont {S.}~\bibnamefont
  {R{\"o}{\ss}ler}}, \bibinfo {author} {\bibfnamefont {T.-H.}\ \bibnamefont
  {Jang}}, \bibinfo {author} {\bibfnamefont {D.~J.}\ \bibnamefont {Kim}},
  \bibinfo {author} {\bibfnamefont {L.~H.}\ \bibnamefont {Tjeng}}, \bibinfo
  {author} {\bibfnamefont {Z.}~\bibnamefont {Fisk}}, \bibinfo {author}
  {\bibfnamefont {F.}~\bibnamefont {Steglich}}, \ and\ \bibinfo {author}
  {\bibfnamefont {S.}~\bibnamefont {Wirth}},\ }\href@noop {} {\bibfield
  {journal} {\bibinfo  {journal} {Proc.\ Natl.\ Acad.\ Sci.\ USA}\ }\textbf
  {\bibinfo {volume} {111}},\ \bibinfo {pages} {4798} (\bibinfo {year}
  {2014})}\BibitemShut {NoStop}%
\bibitem [{\citenamefont {Sun}\ \emph {et~al.}(2018)\citenamefont {Sun},
  \citenamefont {Maldonado}, \citenamefont {Paz}, \citenamefont {Inosov},
  \citenamefont {Schnyder}, \citenamefont {Palacios}, \citenamefont
  {Shitsevalova}, \citenamefont {Filipov},\ and\ \citenamefont {Wahl}}]{sun18}%
  \BibitemOpen
  \bibfield  {author} {\bibinfo {author} {\bibfnamefont {Z.}~\bibnamefont
  {Sun}}, \bibinfo {author} {\bibfnamefont {A.}~\bibnamefont {Maldonado}},
  \bibinfo {author} {\bibfnamefont {W.~S.}\ \bibnamefont {Paz}}, \bibinfo
  {author} {\bibfnamefont {D.~S.}\ \bibnamefont {Inosov}}, \bibinfo {author}
  {\bibfnamefont {A.~P.}\ \bibnamefont {Schnyder}}, \bibinfo {author}
  {\bibfnamefont {J.~J.}\ \bibnamefont {Palacios}}, \bibinfo {author}
  {\bibfnamefont {N.~Y.}\ \bibnamefont {Shitsevalova}}, \bibinfo {author}
  {\bibfnamefont {V.~B.}\ \bibnamefont {Filipov}}, \ and\ \bibinfo {author}
  {\bibfnamefont {P.}~\bibnamefont {Wahl}},\ }\href@noop {} {\bibfield
  {journal} {\bibinfo  {journal} {Phys.\ Rev.\ B}\ }\textbf {\bibinfo {volume}
  {97}},\ \bibinfo {pages} {235107} (\bibinfo {year} {2018})}\BibitemShut
  {NoStop}%
\bibitem [{\citenamefont {Pirie}\ \emph {et~al.}(2020)\citenamefont {Pirie},
  \citenamefont {Liu}, \citenamefont {Soumyanarayanan}, \citenamefont {Chen},
  \citenamefont {He}, \citenamefont {Yee}, \citenamefont {Rosa}, \citenamefont
  {Thompson}, \citenamefont {Kim}, \citenamefont {Fisk}, \citenamefont {Wang},
  \citenamefont {Paglione}, \citenamefont {Morr}, \citenamefont {Hamidian},\
  and\ \citenamefont {Hoffman}}]{pir19}%
  \BibitemOpen
  \bibfield  {author} {\bibinfo {author} {\bibfnamefont {H.}~\bibnamefont
  {Pirie}}, \bibinfo {author} {\bibfnamefont {Y.}~\bibnamefont {Liu}}, \bibinfo
  {author} {\bibfnamefont {A.}~\bibnamefont {Soumyanarayanan}}, \bibinfo
  {author} {\bibfnamefont {P.}~\bibnamefont {Chen}}, \bibinfo {author}
  {\bibfnamefont {Y.}~\bibnamefont {He}}, \bibinfo {author} {\bibfnamefont
  {M.~M.}\ \bibnamefont {Yee}}, \bibinfo {author} {\bibfnamefont {P.~F.~S.}\
  \bibnamefont {Rosa}}, \bibinfo {author} {\bibfnamefont {J.~D.}\ \bibnamefont
  {Thompson}}, \bibinfo {author} {\bibfnamefont {D.-J.}\ \bibnamefont {Kim}},
  \bibinfo {author} {\bibfnamefont {Z.}~\bibnamefont {Fisk}}, \bibinfo {author}
  {\bibfnamefont {X.}~\bibnamefont {Wang}}, \bibinfo {author} {\bibfnamefont
  {J.}~\bibnamefont {Paglione}}, \bibinfo {author} {\bibfnamefont {D.~K.}\
  \bibnamefont {Morr}}, \bibinfo {author} {\bibfnamefont {M.~H.}\ \bibnamefont
  {Hamidian}}, \ and\ \bibinfo {author} {\bibfnamefont {J.~E.}\ \bibnamefont
  {Hoffman}},\ }\href@noop {} {\bibfield  {journal} {\bibinfo  {journal}
  {Nature Phys.}\ }\textbf {\bibinfo {volume} {16}},\ \bibinfo {pages} {52}
  (\bibinfo {year} {2020})}\BibitemShut {NoStop}%
\bibitem [{\citenamefont {R{\"o}{\ss}ler}\ \emph {et~al.}(2016)\citenamefont
  {R{\"o}{\ss}ler}, \citenamefont {Jiao}, \citenamefont {Kim}, \citenamefont
  {Seiro}, \citenamefont {Rasim}, \citenamefont {Steglich}, \citenamefont
  {Tjeng}, \citenamefont {Fisk},\ and\ \citenamefont {Wirth}}]{roe16}%
  \BibitemOpen
  \bibfield  {author} {\bibinfo {author} {\bibfnamefont {S.}~\bibnamefont
  {R{\"o}{\ss}ler}}, \bibinfo {author} {\bibfnamefont {L.}~\bibnamefont
  {Jiao}}, \bibinfo {author} {\bibfnamefont {D.~J.}\ \bibnamefont {Kim}},
  \bibinfo {author} {\bibfnamefont {S.}~\bibnamefont {Seiro}}, \bibinfo
  {author} {\bibfnamefont {K.}~\bibnamefont {Rasim}}, \bibinfo {author}
  {\bibfnamefont {F.}~\bibnamefont {Steglich}}, \bibinfo {author}
  {\bibfnamefont {L.~H.}\ \bibnamefont {Tjeng}}, \bibinfo {author}
  {\bibfnamefont {Z.}~\bibnamefont {Fisk}}, \ and\ \bibinfo {author}
  {\bibfnamefont {S.}~\bibnamefont {Wirth}},\ }\href@noop {} {\bibfield
  {journal} {\bibinfo  {journal} {Philos.\ Mag.}\ }\textbf {\bibinfo {volume}
  {96}},\ \bibinfo {pages} {3262} (\bibinfo {year} {2016})}\BibitemShut
  {NoStop}%
\bibitem [{\citenamefont {Herrmann}\ \emph {et~al.}(2020)\citenamefont
  {Herrmann}, \citenamefont {Hlawenka}, \citenamefont {Siemensmeyer},
  \citenamefont {Weschke}, \citenamefont {S\'{a}nchez-Barriga}, \citenamefont
  {Varykhalov}, \citenamefont {Shitsevalova}, \citenamefont {Dukhnenko},
  \citenamefont {Filipov}, \citenamefont {Gab\'{a}ni}, \citenamefont
  {Flachbart}, \citenamefont {Rader}, \citenamefont {Sterrer},\ and\
  \citenamefont {Rienks}}]{her18}%
  \BibitemOpen
  \bibfield  {author} {\bibinfo {author} {\bibfnamefont {H.}~\bibnamefont
  {Herrmann}}, \bibinfo {author} {\bibfnamefont {P.}~\bibnamefont {Hlawenka}},
  \bibinfo {author} {\bibfnamefont {K.}~\bibnamefont {Siemensmeyer}}, \bibinfo
  {author} {\bibfnamefont {E.}~\bibnamefont {Weschke}}, \bibinfo {author}
  {\bibfnamefont {J.}~\bibnamefont {S\'{a}nchez-Barriga}}, \bibinfo {author}
  {\bibfnamefont {A.}~\bibnamefont {Varykhalov}}, \bibinfo {author}
  {\bibfnamefont {N.~Y.}\ \bibnamefont {Shitsevalova}}, \bibinfo {author}
  {\bibfnamefont {A.~V.}\ \bibnamefont {Dukhnenko}}, \bibinfo {author}
  {\bibfnamefont {V.~B.}\ \bibnamefont {Filipov}}, \bibinfo {author}
  {\bibfnamefont {S.}~\bibnamefont {Gab\'{a}ni}}, \bibinfo {author}
  {\bibfnamefont {K.}~\bibnamefont {Flachbart}}, \bibinfo {author}
  {\bibfnamefont {O.}~\bibnamefont {Rader}}, \bibinfo {author} {\bibfnamefont
  {M.}~\bibnamefont {Sterrer}}, \ and\ \bibinfo {author} {\bibfnamefont
  {E.~D.~L.}\ \bibnamefont {Rienks}},\ }\href@noop {} {\bibfield  {journal}
  {\bibinfo  {journal} {Adv.\ Mater.}\ }\textbf {\bibinfo {volume} {32}}
  (\bibinfo {year} {2020})},\ \bibinfo {note}
  {doi:10.1002/adma.201906725}\BibitemShut {NoStop}%
\bibitem [{\citenamefont {Zhang}\ \emph {et~al.}(2008)\citenamefont {Zhang},
  \citenamefont {{von Moln{\'a}r}}, \citenamefont {Fisk},\ and\ \citenamefont
  {Xiong}}]{zha08}%
  \BibitemOpen
  \bibfield  {author} {\bibinfo {author} {\bibfnamefont {X.}~\bibnamefont
  {Zhang}}, \bibinfo {author} {\bibfnamefont {S.}~\bibnamefont {{von
  Moln{\'a}r}}}, \bibinfo {author} {\bibfnamefont {Z.}~\bibnamefont {Fisk}}, \
  and\ \bibinfo {author} {\bibfnamefont {P.}~\bibnamefont {Xiong}},\
  }\href@noop {} {\bibfield  {journal} {\bibinfo  {journal} {Phys.\ Rev.\
  Lett.}\ }\textbf {\bibinfo {volume} {100}},\ \bibinfo {pages} {167001}
  (\bibinfo {year} {2008})}\BibitemShut {NoStop}%
\bibitem [{\citenamefont {S{\"u}llow}\ \emph {et~al.}(1998)\citenamefont
  {S{\"u}llow}, \citenamefont {Prasad}, \citenamefont {Aronson}, \citenamefont
  {Sarrao}, \citenamefont {Fisk}, \citenamefont {Hristova}, \citenamefont
  {Lacerda}, \citenamefont {Hundley}, \citenamefont {Vigliante},\ and\
  \citenamefont {Gibbs}}]{sue98}%
  \BibitemOpen
  \bibfield  {author} {\bibinfo {author} {\bibfnamefont {S.}~\bibnamefont
  {S{\"u}llow}}, \bibinfo {author} {\bibfnamefont {I.}~\bibnamefont {Prasad}},
  \bibinfo {author} {\bibfnamefont {M.~C.}\ \bibnamefont {Aronson}}, \bibinfo
  {author} {\bibfnamefont {J.~L.}\ \bibnamefont {Sarrao}}, \bibinfo {author}
  {\bibfnamefont {Z.}~\bibnamefont {Fisk}}, \bibinfo {author} {\bibfnamefont
  {D.}~\bibnamefont {Hristova}}, \bibinfo {author} {\bibfnamefont {A.~H.}\
  \bibnamefont {Lacerda}}, \bibinfo {author} {\bibfnamefont {M.~F.}\
  \bibnamefont {Hundley}}, \bibinfo {author} {\bibfnamefont {A.}~\bibnamefont
  {Vigliante}}, \ and\ \bibinfo {author} {\bibfnamefont {D.}~\bibnamefont
  {Gibbs}},\ }\href@noop {} {\bibfield  {journal} {\bibinfo  {journal} {Phys.\
  Rev.\ B}\ }\textbf {\bibinfo {volume} {57}},\ \bibinfo {pages} {5860}
  (\bibinfo {year} {1998})}\BibitemShut {NoStop}%
\bibitem [{\citenamefont {Pohlit}\ \emph {et~al.}(2018)\citenamefont {Pohlit},
  \citenamefont {R{\"o}{\ss}ler}, \citenamefont {Ohno}, \citenamefont {Ohno},
  \citenamefont {{von Moln{\'a}r}}, \citenamefont {Fisk}, \citenamefont
  {M{\"u}ller},\ and\ \citenamefont {Wirth}}]{poh18}%
  \BibitemOpen
  \bibfield  {author} {\bibinfo {author} {\bibfnamefont {M.}~\bibnamefont
  {Pohlit}}, \bibinfo {author} {\bibfnamefont {S.}~\bibnamefont
  {R{\"o}{\ss}ler}}, \bibinfo {author} {\bibfnamefont {Y.}~\bibnamefont
  {Ohno}}, \bibinfo {author} {\bibfnamefont {H.}~\bibnamefont {Ohno}}, \bibinfo
  {author} {\bibfnamefont {S.}~\bibnamefont {{von Moln{\'a}r}}}, \bibinfo
  {author} {\bibfnamefont {Z.}~\bibnamefont {Fisk}}, \bibinfo {author}
  {\bibfnamefont {J.}~\bibnamefont {M{\"u}ller}}, \ and\ \bibinfo {author}
  {\bibfnamefont {S.}~\bibnamefont {Wirth}},\ }\href@noop {} {\bibfield
  {journal} {\bibinfo  {journal} {Phys.\ Rev.\ Lett.}\ }\textbf {\bibinfo
  {volume} {120}},\ \bibinfo {pages} {257201} (\bibinfo {year}
  {2018})}\BibitemShut {NoStop}%
\bibitem [{\citenamefont {Ramankutty}\ \emph {et~al.}(2016)\citenamefont
  {Ramankutty}, \citenamefont {de~Jong}, \citenamefont {Huang}, \citenamefont
  {Zwartsenberg}, \citenamefont {Massee}, \citenamefont {Bay}, \citenamefont
  {Golden},\ and\ \citenamefont {Frantzeskakis}}]{ram16}%
  \BibitemOpen
  \bibfield  {author} {\bibinfo {author} {\bibfnamefont {S.~V.}\ \bibnamefont
  {Ramankutty}}, \bibinfo {author} {\bibfnamefont {N.}~\bibnamefont {de~Jong}},
  \bibinfo {author} {\bibfnamefont {Y.~K.}\ \bibnamefont {Huang}}, \bibinfo
  {author} {\bibfnamefont {B.}~\bibnamefont {Zwartsenberg}}, \bibinfo {author}
  {\bibfnamefont {F.}~\bibnamefont {Massee}}, \bibinfo {author} {\bibfnamefont
  {T.~V.}\ \bibnamefont {Bay}}, \bibinfo {author} {\bibfnamefont {M.~S.}\
  \bibnamefont {Golden}}, \ and\ \bibinfo {author} {\bibfnamefont
  {E.}~\bibnamefont {Frantzeskakis}},\ }\href@noop {} {\bibfield  {journal}
  {\bibinfo  {journal} {J. Electron Spectrosc.\ Relat.\ Phenom.}\ }\textbf
  {\bibinfo {volume} {208}},\ \bibinfo {pages} {43} (\bibinfo {year}
  {2016})}\BibitemShut {NoStop}%
\bibitem [{\citenamefont {Fisk}\ \emph {et~al.}(1979)\citenamefont {Fisk},
  \citenamefont {Johnston}, \citenamefont {Cornut}, \citenamefont {{von
  Moln{\'a}r}}, \citenamefont {Oseroff},\ and\ \citenamefont {Calvo}}]{fis79}%
  \BibitemOpen
  \bibfield  {author} {\bibinfo {author} {\bibfnamefont {Z.}~\bibnamefont
  {Fisk}}, \bibinfo {author} {\bibfnamefont {D.}~\bibnamefont {Johnston}},
  \bibinfo {author} {\bibfnamefont {B.}~\bibnamefont {Cornut}}, \bibinfo
  {author} {\bibfnamefont {S.}~\bibnamefont {{von Moln{\'a}r}}}, \bibinfo
  {author} {\bibfnamefont {S.}~\bibnamefont {Oseroff}}, \ and\ \bibinfo
  {author} {\bibfnamefont {R.}~\bibnamefont {Calvo}},\ }\href@noop {}
  {\bibfield  {journal} {\bibinfo  {journal} {J.\ Appl.\ Phys.}\ }\textbf
  {\bibinfo {volume} {50}},\ \bibinfo {pages} {1911} (\bibinfo {year}
  {1979})}\BibitemShut {NoStop}%
\bibitem [{\citenamefont {Kim}\ \emph {et~al.}(2014{\natexlab{b}})\citenamefont
  {Kim}, \citenamefont {Xia},\ and\ \citenamefont {Fisk}}]{kim14}%
  \BibitemOpen
  \bibfield  {author} {\bibinfo {author} {\bibfnamefont {D.~J.}\ \bibnamefont
  {Kim}}, \bibinfo {author} {\bibfnamefont {J.}~\bibnamefont {Xia}}, \ and\
  \bibinfo {author} {\bibfnamefont {Z.}~\bibnamefont {Fisk}},\ }\href@noop {}
  {\bibfield  {journal} {\bibinfo  {journal} {Nature Mater.}\ }\textbf
  {\bibinfo {volume} {13}},\ \bibinfo {pages} {466} (\bibinfo {year}
  {2014}{\natexlab{b}})}\BibitemShut {NoStop}%
\bibitem [{\citenamefont {Rosa}\ and\ \citenamefont {Fisk}(2018)}]{rosa18}%
  \BibitemOpen
  \bibfield  {author} {\bibinfo {author} {\bibfnamefont {P.~F.~S.}\
  \bibnamefont {Rosa}}\ and\ \bibinfo {author} {\bibfnamefont {Z.}~\bibnamefont
  {Fisk}},\ }in\ \href@noop {} {\emph {\bibinfo {booktitle} {Crystal Growth of
  Intermetallics}}},\ \bibinfo {editor} {edited by\ \bibinfo {editor}
  {\bibfnamefont {P.}~\bibnamefont {Gille}}\ and\ \bibinfo {editor}
  {\bibfnamefont {Y.}~\bibnamefont {Grin}}}\ (\bibinfo  {publisher} {Berlin,
  Boston: De Gruyter},\ \bibinfo {year} {2018})\ pp.\ \bibinfo {pages}
  {49--60}\BibitemShut {NoStop}%
\bibitem [{\citenamefont {{Omicron Nanotechnology GmbH}}()}]{omi}%
  \BibitemOpen
  \bibfield  {author} {\bibinfo {author} {\bibnamefont {{Omicron Nanotechnology
  GmbH}}},\ }\href@noop {} {}\bibinfo {note} {{T}aunusstein
  (Germany)}\BibitemShut {NoStop}%
\bibitem [{\citenamefont {Schubert}\ \emph {et~al.}(2012)\citenamefont
  {Schubert}, \citenamefont {Fehske}, \citenamefont {Fritz},\ and\
  \citenamefont {Vojta}}]{schu12}%
  \BibitemOpen
  \bibfield  {author} {\bibinfo {author} {\bibfnamefont {G.}~\bibnamefont
  {Schubert}}, \bibinfo {author} {\bibfnamefont {H.}~\bibnamefont {Fehske}},
  \bibinfo {author} {\bibfnamefont {L.}~\bibnamefont {Fritz}}, \ and\ \bibinfo
  {author} {\bibfnamefont {M.}~\bibnamefont {Vojta}},\ }\href@noop {}
  {\bibfield  {journal} {\bibinfo  {journal} {Phys.\ Rev.\ B}\ }\textbf
  {\bibinfo {volume} {85}},\ \bibinfo {pages} {201105(R)} (\bibinfo {year}
  {2012})}\BibitemShut {NoStop}%
\bibitem [{\citenamefont {Wolgast}\ \emph {et~al.}(2015)\citenamefont
  {Wolgast}, \citenamefont {Eo}, \citenamefont {{\"O}zt{\"u}rk}, \citenamefont
  {Li}, \citenamefont {Xiang}, \citenamefont {Tinsman}, \citenamefont {Asaba},
  \citenamefont {Lawson}, \citenamefont {Yu}, \citenamefont {Allen},
  \citenamefont {Sun}, \citenamefont {Li}, \citenamefont {Kurdak},
  \citenamefont {Kim},\ and\ \citenamefont {Fisk}}]{wol15}%
  \BibitemOpen
  \bibfield  {author} {\bibinfo {author} {\bibfnamefont {S.}~\bibnamefont
  {Wolgast}}, \bibinfo {author} {\bibfnamefont {Y.~S.}\ \bibnamefont {Eo}},
  \bibinfo {author} {\bibfnamefont {T.}~\bibnamefont {{\"O}zt{\"u}rk}},
  \bibinfo {author} {\bibfnamefont {G.}~\bibnamefont {Li}}, \bibinfo {author}
  {\bibfnamefont {Z.}~\bibnamefont {Xiang}}, \bibinfo {author} {\bibfnamefont
  {C.}~\bibnamefont {Tinsman}}, \bibinfo {author} {\bibfnamefont
  {T.}~\bibnamefont {Asaba}}, \bibinfo {author} {\bibfnamefont
  {B.}~\bibnamefont {Lawson}}, \bibinfo {author} {\bibfnamefont
  {F.}~\bibnamefont {Yu}}, \bibinfo {author} {\bibfnamefont {J.~W.}\
  \bibnamefont {Allen}}, \bibinfo {author} {\bibfnamefont {K.}~\bibnamefont
  {Sun}}, \bibinfo {author} {\bibfnamefont {L.}~\bibnamefont {Li}}, \bibinfo
  {author} {\bibfnamefont {C.}~\bibnamefont {Kurdak}}, \bibinfo {author}
  {\bibfnamefont {D.-J.}\ \bibnamefont {Kim}}, \ and\ \bibinfo {author}
  {\bibfnamefont {Z.}~\bibnamefont {Fisk}},\ }\href@noop {} {\bibfield
  {journal} {\bibinfo  {journal} {Phys.\ Rev.\ B}\ }\textbf {\bibinfo {volume}
  {92}},\ \bibinfo {pages} {115110} (\bibinfo {year} {2015})}\BibitemShut
  {NoStop}%
\bibitem [{\citenamefont {Yee}\ \emph {et~al.}(2013)\citenamefont {Yee},
  \citenamefont {He}, \citenamefont {Soumyanarayanan}, \citenamefont {Kim},
  \citenamefont {Fisk},\ and\ \citenamefont {Hoffman}}]{yee13}%
  \BibitemOpen
  \bibfield  {author} {\bibinfo {author} {\bibfnamefont {M.~M.}\ \bibnamefont
  {Yee}}, \bibinfo {author} {\bibfnamefont {Y.}~\bibnamefont {He}}, \bibinfo
  {author} {\bibfnamefont {A.}~\bibnamefont {Soumyanarayanan}}, \bibinfo
  {author} {\bibfnamefont {D.-J.}\ \bibnamefont {Kim}}, \bibinfo {author}
  {\bibfnamefont {Z.}~\bibnamefont {Fisk}}, \ and\ \bibinfo {author}
  {\bibfnamefont {J.~E.}\ \bibnamefont {Hoffman}},\ }\href@noop {} {\enquote
  {\bibinfo {title} {Imaging the {K}ondo insulating gap on {S}m{B}$_6$.}}\ }
  (\bibinfo {year} {2013}),\ \bibinfo {note} {{a}rXiv:1308.1085}\BibitemShut
  {NoStop}%
\bibitem [{\citenamefont {Miyazaki}\ \emph {et~al.}(2012)\citenamefont
  {Miyazaki}, \citenamefont {Hajiri}, \citenamefont {Ito}, \citenamefont
  {Kunii},\ and\ \citenamefont {Kimura}}]{miy12}%
  \BibitemOpen
  \bibfield  {author} {\bibinfo {author} {\bibfnamefont {H.}~\bibnamefont
  {Miyazaki}}, \bibinfo {author} {\bibfnamefont {T.}~\bibnamefont {Hajiri}},
  \bibinfo {author} {\bibfnamefont {T.}~\bibnamefont {Ito}}, \bibinfo {author}
  {\bibfnamefont {S.}~\bibnamefont {Kunii}}, \ and\ \bibinfo {author}
  {\bibfnamefont {S.~I.}\ \bibnamefont {Kimura}},\ }\href@noop {} {\bibfield
  {journal} {\bibinfo  {journal} {Phys.\ Rev.\ B}\ }\textbf {\bibinfo {volume}
  {86}},\ \bibinfo {pages} {075105} (\bibinfo {year} {2012})}\BibitemShut
  {NoStop}%
\bibitem [{\citenamefont {Miyamachi}\ \emph {et~al.}(2017)\citenamefont
  {Miyamachi}, \citenamefont {Suga}, \citenamefont {Ellguth}, \citenamefont
  {Tusche}, \citenamefont {Schneider}, \citenamefont {Iga},\ and\ \citenamefont
  {Komori}}]{miy17}%
  \BibitemOpen
  \bibfield  {author} {\bibinfo {author} {\bibfnamefont {T.}~\bibnamefont
  {Miyamachi}}, \bibinfo {author} {\bibfnamefont {S.}~\bibnamefont {Suga}},
  \bibinfo {author} {\bibfnamefont {M.}~\bibnamefont {Ellguth}}, \bibinfo
  {author} {\bibfnamefont {C.}~\bibnamefont {Tusche}}, \bibinfo {author}
  {\bibfnamefont {C.~M.}\ \bibnamefont {Schneider}}, \bibinfo {author}
  {\bibfnamefont {F.}~\bibnamefont {Iga}}, \ and\ \bibinfo {author}
  {\bibfnamefont {F.}~\bibnamefont {Komori}},\ }\href@noop {} {\bibfield
  {journal} {\bibinfo  {journal} {Sci.\ Rep.}\ }\textbf {\bibinfo {volume}
  {7}},\ \bibinfo {pages} {12837} (\bibinfo {year} {2017})}\BibitemShut
  {NoStop}%
\bibitem [{\citenamefont {Buchsteiner}\ \emph {et~al.}(2019)\citenamefont
  {Buchsteiner}, \citenamefont {Sohn}, \citenamefont {Horstmann}, \citenamefont
  {Voigt}, \citenamefont {Hatnean}, \citenamefont {Balakrishnan}, \citenamefont
  {Ropers}, \citenamefont {Bl{\"o}chl},\ and\ \citenamefont
  {Wenderoth}}]{buc19}%
  \BibitemOpen
  \bibfield  {author} {\bibinfo {author} {\bibfnamefont {P.}~\bibnamefont
  {Buchsteiner}}, \bibinfo {author} {\bibfnamefont {F.}~\bibnamefont {Sohn}},
  \bibinfo {author} {\bibfnamefont {J.~G.}\ \bibnamefont {Horstmann}}, \bibinfo
  {author} {\bibfnamefont {J.}~\bibnamefont {Voigt}}, \bibinfo {author}
  {\bibfnamefont {M.~C.}\ \bibnamefont {Hatnean}}, \bibinfo {author}
  {\bibfnamefont {G.}~\bibnamefont {Balakrishnan}}, \bibinfo {author}
  {\bibfnamefont {C.}~\bibnamefont {Ropers}}, \bibinfo {author} {\bibfnamefont
  {P.~E.}\ \bibnamefont {Bl{\"o}chl}}, \ and\ \bibinfo {author} {\bibfnamefont
  {M.}~\bibnamefont {Wenderoth}},\ }\href@noop {} {\bibfield  {journal}
  {\bibinfo  {journal} {Phys.\ Rev.\ B}\ }\textbf {\bibinfo {volume} {100}},\
  \bibinfo {pages} {205407} (\bibinfo {year} {2019})}\BibitemShut {NoStop}%
\bibitem [{\citenamefont {Yoo}\ and\ \citenamefont {Weitering}(2002)}]{yoo02}%
  \BibitemOpen
  \bibfield  {author} {\bibinfo {author} {\bibfnamefont {K.}~\bibnamefont
  {Yoo}}\ and\ \bibinfo {author} {\bibfnamefont {H.~H.}\ \bibnamefont
  {Weitering}},\ }\href@noop {} {\bibfield  {journal} {\bibinfo  {journal}
  {Phys.\ Rev.\ B}\ }\textbf {\bibinfo {volume} {65}},\ \bibinfo {pages}
  {115424} (\bibinfo {year} {2002})}\BibitemShut {NoStop}%
\bibitem [{\citenamefont {Jiao}\ \emph {et~al.}(2016)\citenamefont {Jiao},
  \citenamefont {R{\"o}{\ss}ler}, \citenamefont {Kim}, \citenamefont {Tjeng},
  \citenamefont {Fisk}, \citenamefont {Steglich},\ and\ \citenamefont
  {Wirth}}]{jiao16}%
  \BibitemOpen
  \bibfield  {author} {\bibinfo {author} {\bibfnamefont {L.}~\bibnamefont
  {Jiao}}, \bibinfo {author} {\bibfnamefont {S.}~\bibnamefont
  {R{\"o}{\ss}ler}}, \bibinfo {author} {\bibfnamefont {D.~J.}\ \bibnamefont
  {Kim}}, \bibinfo {author} {\bibfnamefont {L.~H.}\ \bibnamefont {Tjeng}},
  \bibinfo {author} {\bibfnamefont {Z.}~\bibnamefont {Fisk}}, \bibinfo {author}
  {\bibfnamefont {F.}~\bibnamefont {Steglich}}, \ and\ \bibinfo {author}
  {\bibfnamefont {S.}~\bibnamefont {Wirth}},\ }\href@noop {} {\bibfield
  {journal} {\bibinfo  {journal} {Nature Commun.}\ }\textbf {\bibinfo {volume}
  {7}},\ \bibinfo {pages} {13762} (\bibinfo {year} {2016})}\BibitemShut
  {NoStop}%
\bibitem [{\citenamefont {Matt}\ \emph {et~al.}(2018)\citenamefont {Matt},
  \citenamefont {Pirie}, \citenamefont {Soumyanarayanan}, \citenamefont {Yee},
  \citenamefont {He}, \citenamefont {Larson}, \citenamefont {Paz},
  \citenamefont {Palacios}, \citenamefont {Hamidian},\ and\ \citenamefont
  {Hoffman}}]{mat18}%
  \BibitemOpen
  \bibfield  {author} {\bibinfo {author} {\bibfnamefont {C.~E.}\ \bibnamefont
  {Matt}}, \bibinfo {author} {\bibfnamefont {H.}~\bibnamefont {Pirie}},
  \bibinfo {author} {\bibfnamefont {A.}~\bibnamefont {Soumyanarayanan}},
  \bibinfo {author} {\bibfnamefont {M.~M.}\ \bibnamefont {Yee}}, \bibinfo
  {author} {\bibfnamefont {Y.}~\bibnamefont {He}}, \bibinfo {author}
  {\bibfnamefont {D.~T.}\ \bibnamefont {Larson}}, \bibinfo {author}
  {\bibfnamefont {W.~S.}\ \bibnamefont {Paz}}, \bibinfo {author} {\bibfnamefont
  {J.~J.}\ \bibnamefont {Palacios}}, \bibinfo {author} {\bibfnamefont {M.~H.}\
  \bibnamefont {Hamidian}}, \ and\ \bibinfo {author} {\bibfnamefont {J.~E.}\
  \bibnamefont {Hoffman}},\ }\href@noop {} {\enquote {\bibinfo {title}
  {Consistency between {ARPES} and {STM} measurements on {S}m{B}$_6$.}}\ }
  (\bibinfo {year} {2018}),\ \bibinfo {note} {arXiv:1810.13442}\BibitemShut
  {NoStop}%
\bibitem [{\citenamefont {Zhang}\ \emph {et~al.}(2013)\citenamefont {Zhang},
  \citenamefont {Butch}, \citenamefont {Syers}, \citenamefont {Ziemak},
  \citenamefont {Greene},\ and\ \citenamefont {Paglione}}]{zha13}%
  \BibitemOpen
  \bibfield  {author} {\bibinfo {author} {\bibfnamefont {X.}~\bibnamefont
  {Zhang}}, \bibinfo {author} {\bibfnamefont {N.~P.}\ \bibnamefont {Butch}},
  \bibinfo {author} {\bibfnamefont {P.}~\bibnamefont {Syers}}, \bibinfo
  {author} {\bibfnamefont {S.}~\bibnamefont {Ziemak}}, \bibinfo {author}
  {\bibfnamefont {R.~L.}\ \bibnamefont {Greene}}, \ and\ \bibinfo {author}
  {\bibfnamefont {J.}~\bibnamefont {Paglione}},\ }\href@noop {} {\bibfield
  {journal} {\bibinfo  {journal} {Phys.\ Rev.\ X}\ }\textbf {\bibinfo {volume}
  {3}},\ \bibinfo {pages} {011011} (\bibinfo {year} {2013})}\BibitemShut
  {NoStop}%
\bibitem [{\citenamefont {Xu}\ \emph {et~al.}(2013)\citenamefont {Xu},
  \citenamefont {Shi}, \citenamefont {Biswas}, \citenamefont {Matt},
  \citenamefont {Dhaka}, \citenamefont {Huang}, \citenamefont {Plumb},
  \citenamefont {Radovi{\'c}}, \citenamefont {Dil}, \citenamefont
  {Pomjakushina}, \citenamefont {Conder}, \citenamefont {Amato}, \citenamefont
  {Salman}, \citenamefont {Paul}, \citenamefont {Mesot}, \citenamefont {Ding},\
  and\ \citenamefont {Shi}}]{nxu13}%
  \BibitemOpen
  \bibfield  {author} {\bibinfo {author} {\bibfnamefont {N.}~\bibnamefont
  {Xu}}, \bibinfo {author} {\bibfnamefont {X.}~\bibnamefont {Shi}}, \bibinfo
  {author} {\bibfnamefont {P.~K.}\ \bibnamefont {Biswas}}, \bibinfo {author}
  {\bibfnamefont {C.~E.}\ \bibnamefont {Matt}}, \bibinfo {author}
  {\bibfnamefont {R.~S.}\ \bibnamefont {Dhaka}}, \bibinfo {author}
  {\bibfnamefont {Y.}~\bibnamefont {Huang}}, \bibinfo {author} {\bibfnamefont
  {N.~C.}\ \bibnamefont {Plumb}}, \bibinfo {author} {\bibfnamefont
  {M.}~\bibnamefont {Radovi{\'c}}}, \bibinfo {author} {\bibfnamefont {J.~H.}\
  \bibnamefont {Dil}}, \bibinfo {author} {\bibfnamefont {E.}~\bibnamefont
  {Pomjakushina}}, \bibinfo {author} {\bibfnamefont {K.}~\bibnamefont
  {Conder}}, \bibinfo {author} {\bibfnamefont {A.}~\bibnamefont {Amato}},
  \bibinfo {author} {\bibfnamefont {Z.}~\bibnamefont {Salman}}, \bibinfo
  {author} {\bibfnamefont {D.~M.}\ \bibnamefont {Paul}}, \bibinfo {author}
  {\bibfnamefont {J.}~\bibnamefont {Mesot}}, \bibinfo {author} {\bibfnamefont
  {H.}~\bibnamefont {Ding}}, \ and\ \bibinfo {author} {\bibfnamefont
  {M.}~\bibnamefont {Shi}},\ }\href@noop {} {\bibfield  {journal} {\bibinfo
  {journal} {Phys.\ Rev.\ B}\ }\textbf {\bibinfo {volume} {88}},\ \bibinfo
  {pages} {121102} (\bibinfo {year} {2013})}\BibitemShut {NoStop}%
\bibitem [{\citenamefont {Frantzeskakis}\ \emph {et~al.}(2013)\citenamefont
  {Frantzeskakis}, \citenamefont {{de Jong}}, \citenamefont {Zwartsenberg},
  \citenamefont {Huang}, \citenamefont {Pan}, \citenamefont {Zhang},
  \citenamefont {Zhang}, \citenamefont {Zhang}, \citenamefont {Bao},
  \citenamefont {Tegus}, \citenamefont {Varykhalov}, \citenamefont {{de
  Visser}},\ and\ \citenamefont {Golden}}]{fra13}%
  \BibitemOpen
  \bibfield  {author} {\bibinfo {author} {\bibfnamefont {E.}~\bibnamefont
  {Frantzeskakis}}, \bibinfo {author} {\bibfnamefont {N.}~\bibnamefont {{de
  Jong}}}, \bibinfo {author} {\bibfnamefont {B.}~\bibnamefont {Zwartsenberg}},
  \bibinfo {author} {\bibfnamefont {Y.~K.}\ \bibnamefont {Huang}}, \bibinfo
  {author} {\bibfnamefont {Y.}~\bibnamefont {Pan}}, \bibinfo {author}
  {\bibfnamefont {X.}~\bibnamefont {Zhang}}, \bibinfo {author} {\bibfnamefont
  {J.~X.}\ \bibnamefont {Zhang}}, \bibinfo {author} {\bibfnamefont {F.~X.}\
  \bibnamefont {Zhang}}, \bibinfo {author} {\bibfnamefont {L.~H.}\ \bibnamefont
  {Bao}}, \bibinfo {author} {\bibfnamefont {O.}~\bibnamefont {Tegus}}, \bibinfo
  {author} {\bibfnamefont {A.}~\bibnamefont {Varykhalov}}, \bibinfo {author}
  {\bibfnamefont {A.}~\bibnamefont {{de Visser}}}, \ and\ \bibinfo {author}
  {\bibfnamefont {M.~S.}\ \bibnamefont {Golden}},\ }\href@noop {} {\bibfield
  {journal} {\bibinfo  {journal} {Phys.\ Rev.\ X}\ }\textbf {\bibinfo {volume}
  {3}},\ \bibinfo {pages} {041024} (\bibinfo {year} {2013})}\BibitemShut
  {NoStop}%
\bibitem [{\citenamefont {Fuhrman}\ \emph {et~al.}(2018)\citenamefont
  {Fuhrman}, \citenamefont {Chamorro}, \citenamefont {Alekseev}, \citenamefont
  {Mignot}, \citenamefont {Keller}, \citenamefont {Rodriguez-Rivera},
  \citenamefont {Qiu}, \citenamefont {Nikoli\'{c}}, \citenamefont {McQueen},\
  and\ \citenamefont {Broholm}}]{fuh17}%
  \BibitemOpen
  \bibfield  {author} {\bibinfo {author} {\bibfnamefont {W.~T.}\ \bibnamefont
  {Fuhrman}}, \bibinfo {author} {\bibfnamefont {J.~R.}\ \bibnamefont
  {Chamorro}}, \bibinfo {author} {\bibfnamefont {P.}~\bibnamefont {Alekseev}},
  \bibinfo {author} {\bibfnamefont {J.-M.}\ \bibnamefont {Mignot}}, \bibinfo
  {author} {\bibfnamefont {T.}~\bibnamefont {Keller}}, \bibinfo {author}
  {\bibfnamefont {J.~A.}\ \bibnamefont {Rodriguez-Rivera}}, \bibinfo {author}
  {\bibfnamefont {Y.}~\bibnamefont {Qiu}}, \bibinfo {author} {\bibfnamefont
  {P.}~\bibnamefont {Nikoli\'{c}}}, \bibinfo {author} {\bibfnamefont {T.~M.}\
  \bibnamefont {McQueen}}, \ and\ \bibinfo {author} {\bibfnamefont {C.~L.}\
  \bibnamefont {Broholm}},\ }\href@noop {} {\bibfield  {journal} {\bibinfo
  {journal} {Nature Commun.}\ }\textbf {\bibinfo {volume} {9}},\ \bibinfo
  {pages} {1539} (\bibinfo {year} {2018})}\BibitemShut {NoStop}%
\bibitem [{\citenamefont {Valentine}\ \emph {et~al.}(2018)\citenamefont
  {Valentine}, \citenamefont {Koohpayeh}, \citenamefont {Phelan}, \citenamefont
  {McQueen}, \citenamefont {Rosa}, \citenamefont {Fisk},\ and\ \citenamefont
  {Drichko}}]{val18}%
  \BibitemOpen
  \bibfield  {author} {\bibinfo {author} {\bibfnamefont {M.~E.}\ \bibnamefont
  {Valentine}}, \bibinfo {author} {\bibfnamefont {S.}~\bibnamefont
  {Koohpayeh}}, \bibinfo {author} {\bibfnamefont {W.~A.}\ \bibnamefont
  {Phelan}}, \bibinfo {author} {\bibfnamefont {T.~M.}\ \bibnamefont {McQueen}},
  \bibinfo {author} {\bibfnamefont {P.~F.~S.}\ \bibnamefont {Rosa}}, \bibinfo
  {author} {\bibfnamefont {Z.}~\bibnamefont {Fisk}}, \ and\ \bibinfo {author}
  {\bibfnamefont {N.}~\bibnamefont {Drichko}},\ }\href@noop {} {\bibfield
  {journal} {\bibinfo  {journal} {Physica B}\ }\textbf {\bibinfo {volume}
  {536}},\ \bibinfo {pages} {60} (\bibinfo {year} {2018})}\BibitemShut
  {NoStop}%
\bibitem [{\citenamefont {Konovalova}\ \emph {et~al.}(1982)\citenamefont
  {Konovalova}, \citenamefont {Paderno}, \citenamefont {Lundstrem},
  \citenamefont {Finkel'shtein}, \citenamefont {Efremova},\ and\ \citenamefont
  {Dudnik}}]{kon82}%
  \BibitemOpen
  \bibfield  {author} {\bibinfo {author} {\bibfnamefont {E.~S.}\ \bibnamefont
  {Konovalova}}, \bibinfo {author} {\bibfnamefont {Y.~B.}\ \bibnamefont
  {Paderno}}, \bibinfo {author} {\bibfnamefont {T.}~\bibnamefont {Lundstrem}},
  \bibinfo {author} {\bibfnamefont {L.~D.}\ \bibnamefont {Finkel'shtein}},
  \bibinfo {author} {\bibfnamefont {N.~N.}\ \bibnamefont {Efremova}}, \ and\
  \bibinfo {author} {\bibfnamefont {E.~M.}\ \bibnamefont {Dudnik}},\
  }\href@noop {} {\bibfield  {journal} {\bibinfo  {journal} {Powder Metall Met.
  Ceram.}\ }\textbf {\bibinfo {volume} {21}},\ \bibinfo {pages} {820} (\bibinfo
  {year} {1982})}\BibitemShut {NoStop}%
\bibitem [{\citenamefont {Phelan}\ \emph {et~al.}(2016)\citenamefont {Phelan},
  \citenamefont {Koohpayeh}, \citenamefont {Cottingham}, \citenamefont
  {Tutmaher}, \citenamefont {Leiner}, \citenamefont {Lumsden}, \citenamefont
  {Lavelle}, \citenamefont {Wang}, \citenamefont {Hoffmann}, \citenamefont
  {Siegler}, \citenamefont {Haldolaarachchige}, \citenamefont {Young},\ and\
  \citenamefont {McQueen}}]{phe16}%
  \BibitemOpen
  \bibfield  {author} {\bibinfo {author} {\bibfnamefont {W.~A.}\ \bibnamefont
  {Phelan}}, \bibinfo {author} {\bibfnamefont {S.~M.}\ \bibnamefont
  {Koohpayeh}}, \bibinfo {author} {\bibfnamefont {P.}~\bibnamefont
  {Cottingham}}, \bibinfo {author} {\bibfnamefont {J.~A.}\ \bibnamefont
  {Tutmaher}}, \bibinfo {author} {\bibfnamefont {J.~C.}\ \bibnamefont
  {Leiner}}, \bibinfo {author} {\bibfnamefont {M.~D.}\ \bibnamefont {Lumsden}},
  \bibinfo {author} {\bibfnamefont {C.~M.}\ \bibnamefont {Lavelle}}, \bibinfo
  {author} {\bibfnamefont {X.~P.}\ \bibnamefont {Wang}}, \bibinfo {author}
  {\bibfnamefont {C.}~\bibnamefont {Hoffmann}}, \bibinfo {author}
  {\bibfnamefont {M.~A.}\ \bibnamefont {Siegler}}, \bibinfo {author}
  {\bibfnamefont {N.}~\bibnamefont {Haldolaarachchige}}, \bibinfo {author}
  {\bibfnamefont {D.~P.}\ \bibnamefont {Young}}, \ and\ \bibinfo {author}
  {\bibfnamefont {T.~M.}\ \bibnamefont {McQueen}},\ }\href@noop {} {\bibfield
  {journal} {\bibinfo  {journal} {Sci.\ Rep.}\ }\textbf {\bibinfo {volume}
  {6}},\ \bibinfo {pages} {20860} (\bibinfo {year} {2016})}\BibitemShut
  {NoStop}%
\bibitem [{\citenamefont {Thomas}\ \emph {et~al.}(2019)\citenamefont {Thomas},
  \citenamefont {Ding}, \citenamefont {Ronning}, \citenamefont {Zapf},
  \citenamefont {Thompson}, \citenamefont {Fisk}, \citenamefont {Xia},\ and\
  \citenamefont {Rosa}}]{tho19}%
  \BibitemOpen
  \bibfield  {author} {\bibinfo {author} {\bibfnamefont {S.}~\bibnamefont
  {Thomas}}, \bibinfo {author} {\bibfnamefont {X.}~\bibnamefont {Ding}},
  \bibinfo {author} {\bibfnamefont {F.}~\bibnamefont {Ronning}}, \bibinfo
  {author} {\bibfnamefont {V.}~\bibnamefont {Zapf}}, \bibinfo {author}
  {\bibfnamefont {J.}~\bibnamefont {Thompson}}, \bibinfo {author}
  {\bibfnamefont {Z.}~\bibnamefont {Fisk}}, \bibinfo {author} {\bibfnamefont
  {J.}~\bibnamefont {Xia}}, \ and\ \bibinfo {author} {\bibfnamefont
  {P.}~\bibnamefont {Rosa}},\ }\href@noop {} {\bibfield  {journal} {\bibinfo
  {journal} {Phys.\ Rev.\ Lett.}\ }\textbf {\bibinfo {volume} {122}},\ \bibinfo
  {pages} {166401} (\bibinfo {year} {2019})}\BibitemShut {NoStop}%
\bibitem [{\citenamefont {Zandbergen}\ \emph {et~al.}(1994)\citenamefont
  {Zandbergen}, \citenamefont {Gortenmulder}, \citenamefont {Sarrac},
  \citenamefont {Harrison}, \citenamefont {{de Andrade}}, \citenamefont
  {Hermann}, \citenamefont {Han}, \citenamefont {Fisk}, \citenamefont {Maple},\
  and\ \citenamefont {Cava}}]{zan94}%
  \BibitemOpen
  \bibfield  {author} {\bibinfo {author} {\bibfnamefont {H.~W.}\ \bibnamefont
  {Zandbergen}}, \bibinfo {author} {\bibfnamefont {T.~J.}\ \bibnamefont
  {Gortenmulder}}, \bibinfo {author} {\bibfnamefont {J.~L.}\ \bibnamefont
  {Sarrac}}, \bibinfo {author} {\bibfnamefont {J.~C.}\ \bibnamefont
  {Harrison}}, \bibinfo {author} {\bibfnamefont {M.~C.}\ \bibnamefont {{de
  Andrade}}}, \bibinfo {author} {\bibfnamefont {J.}~\bibnamefont {Hermann}},
  \bibinfo {author} {\bibfnamefont {S.~H.}\ \bibnamefont {Han}}, \bibinfo
  {author} {\bibfnamefont {Z.}~\bibnamefont {Fisk}}, \bibinfo {author}
  {\bibfnamefont {M.~B.}\ \bibnamefont {Maple}}, \ and\ \bibinfo {author}
  {\bibfnamefont {R.~J.}\ \bibnamefont {Cava}},\ }\href@noop {} {\bibfield
  {journal} {\bibinfo  {journal} {Physica C}\ }\textbf {\bibinfo {volume}
  {232}},\ \bibinfo {pages} {328} (\bibinfo {year} {1994})}\BibitemShut
  {NoStop}%
\bibitem [{\citenamefont {Ozcomert}\ and\ \citenamefont
  {Trenary}(1992)}]{ozc92}%
  \BibitemOpen
  \bibfield  {author} {\bibinfo {author} {\bibfnamefont {J.~S.}\ \bibnamefont
  {Ozcomert}}\ and\ \bibinfo {author} {\bibfnamefont {M.}~\bibnamefont
  {Trenary}},\ }\href@noop {} {\bibfield  {journal} {\bibinfo  {journal}
  {Surf.\ Sci.\ Lett.}\ }\textbf {\bibinfo {volume} {265}},\ \bibinfo {pages}
  {L227} (\bibinfo {year} {1992})}\BibitemShut {NoStop}%
\bibitem [{\citenamefont {Massidda}\ \emph {et~al.}(1997)\citenamefont
  {Massidda}, \citenamefont {Continenza}, \citenamefont {de~Pascale},\ and\
  \citenamefont {Monnier}}]{mas97}%
  \BibitemOpen
  \bibfield  {author} {\bibinfo {author} {\bibfnamefont {S.}~\bibnamefont
  {Massidda}}, \bibinfo {author} {\bibfnamefont {A.}~\bibnamefont
  {Continenza}}, \bibinfo {author} {\bibfnamefont {T.~M.}\ \bibnamefont
  {de~Pascale}}, \ and\ \bibinfo {author} {\bibfnamefont {R.}~\bibnamefont
  {Monnier}},\ }\href@noop {} {\bibfield  {journal} {\bibinfo  {journal} {Z.\
  Phys.\ B: Condens.\ Matter}\ }\textbf {\bibinfo {volume} {102}},\ \bibinfo
  {pages} {83} (\bibinfo {year} {1997})}\BibitemShut {NoStop}%
\bibitem [{\citenamefont {Jiao}\ \emph {et~al.}(2018)\citenamefont {Jiao},
  \citenamefont {R{\"o}{\ss}ler}, \citenamefont {Kasinathan}, \citenamefont
  {Rosa}, \citenamefont {Guo}, \citenamefont {Yuan}, \citenamefont {Liu},
  \citenamefont {Fisk}, \citenamefont {Steglich},\ and\ \citenamefont
  {Wirth}}]{jiao18}%
  \BibitemOpen
  \bibfield  {author} {\bibinfo {author} {\bibfnamefont {L.}~\bibnamefont
  {Jiao}}, \bibinfo {author} {\bibfnamefont {S.}~\bibnamefont
  {R{\"o}{\ss}ler}}, \bibinfo {author} {\bibfnamefont {D.}~\bibnamefont
  {Kasinathan}}, \bibinfo {author} {\bibfnamefont {P.~F.~S.}\ \bibnamefont
  {Rosa}}, \bibinfo {author} {\bibfnamefont {C.}~\bibnamefont {Guo}}, \bibinfo
  {author} {\bibfnamefont {H.}~\bibnamefont {Yuan}}, \bibinfo {author}
  {\bibfnamefont {C.-X.}\ \bibnamefont {Liu}}, \bibinfo {author} {\bibfnamefont
  {Z.}~\bibnamefont {Fisk}}, \bibinfo {author} {\bibfnamefont {F.}~\bibnamefont
  {Steglich}}, \ and\ \bibinfo {author} {\bibfnamefont {S.}~\bibnamefont
  {Wirth}},\ }\href@noop {} {\bibfield  {journal} {\bibinfo  {journal} {Sci.
  Adv.}\ }\textbf {\bibinfo {volume} {4}},\ \bibinfo {pages} {eaau4886}
  (\bibinfo {year} {2018})}\BibitemShut {NoStop}%
\bibitem [{\citenamefont {Liu}\ \emph {et~al.}(2009)\citenamefont {Liu},
  \citenamefont {Liu}, \citenamefont {Xu}, \citenamefont {Qi},\ and\
  \citenamefont {Zhang}}]{liu09}%
  \BibitemOpen
  \bibfield  {author} {\bibinfo {author} {\bibfnamefont {Q.}~\bibnamefont
  {Liu}}, \bibinfo {author} {\bibfnamefont {C.-X.}\ \bibnamefont {Liu}},
  \bibinfo {author} {\bibfnamefont {C.}~\bibnamefont {Xu}}, \bibinfo {author}
  {\bibfnamefont {X.-L.}\ \bibnamefont {Qi}}, \ and\ \bibinfo {author}
  {\bibfnamefont {S.-C.}\ \bibnamefont {Zhang}},\ }\href@noop {} {\bibfield
  {journal} {\bibinfo  {journal} {Phys.\ Rev.\ Lett.}\ }\textbf {\bibinfo
  {volume} {102}},\ \bibinfo {pages} {156603} (\bibinfo {year}
  {2009})}\BibitemShut {NoStop}%
\bibitem [{\citenamefont {Wang}\ \emph {et~al.}(2010)\citenamefont {Wang},
  \citenamefont {Wang},\ and\ \citenamefont {Zhang}}]{wang10}%
  \BibitemOpen
  \bibfield  {author} {\bibinfo {author} {\bibfnamefont {Q.-H.}\ \bibnamefont
  {Wang}}, \bibinfo {author} {\bibfnamefont {D.}~\bibnamefont {Wang}}, \ and\
  \bibinfo {author} {\bibfnamefont {F.-C.}\ \bibnamefont {Zhang}},\ }\href@noop
  {} {\bibfield  {journal} {\bibinfo  {journal} {Phys.\ Rev.\ B}\ }\textbf
  {\bibinfo {volume} {81}},\ \bibinfo {pages} {035104} (\bibinfo {year}
  {2010})}\BibitemShut {NoStop}%
\end{thebibliography}
\end{document}